%% file: main.tex
\documentclass[aps,prd,11pt,superscriptaddress]{revtex4-2}
\pdfoutput=1

\input{common/preamble}

\usepackage[pdftex,bookmarks,hidelinks]{hyperref}

\begin{document}

\pagestyle{titlepage}
%\cleardoublepage

\input{common/init}

\clearpage

\input{sections/exec-summary}
%\cleardoublepage
\newpage
\input{sections/introduction}
\input{sections/expt_design.tex}
\input{sections/phys_sens.tex}
\input{sections/practical.tex}
\input{sections/snowmass}
\input{sections/conclusions}

\input{sections/acknowledgements}

%\clearpage

\input{common/final}
\end{document}

%% file: common/preamble.tex
\usepackage[utf8]{inputenc}
\usepackage{xspace}

\usepackage{graphicx}

\graphicspath{ {/graphics/} }

\let\OLDthebibliography\thebibliography
\renewcommand\thebibliography[1]{
  \OLDthebibliography{#1}
  \setlength{\parskip}{0pt}
  \setlength{\itemsep}{0pt plus 0.3ex}
}

\newif\ifdp
\newif\ifsp

\input{common/defs.tex}

\input{generated/parameters.tex}

%% file: common/defs.tex
\def\explong{White Paper on New Opportunities at the Next-Generation Neutrino Experiments\xspace}

\newcommand{\numu}{\ensuremath{\nu_\mu}\xspace}
\newcommand{\nue}{\ensuremath{\nu_e}\xspace}

\newcommand{\anumu}{\ensuremath{\bar\nu_\mu}\xspace}
\newcommand{\anue}{\ensuremath{\bar\nu_e}\xspace}

\newcommand{\dm}[1]{\ensuremath{\Delta m^2_{#1}}\xspace} % example: \dm{12}

\newcommand{\sinst}[1]{\ensuremath{\sin^2\theta_{#1}}\xspace} % example \sinst{12}
\newcommand{\sinstt}[1]{\ensuremath{\sin^22\theta_{#1}}\xspace}  % example \sinstt{12}

\newcommand{\deltacp}{\ensuremath{\delta_{\rm CP}}\xspace}   % example \deltacp
\newcommand{\nuxtonux}[2]{\ensuremath{\nu_{#1} \to \nu_{#2}}\xspace}  % example \nuxtonux23 (no {...} )

\newcommand{\numutonue}{\nuxtonux{\mu}{e}}

\newcommand{\nuxtonuxbar}[2]{\ensuremath{\bar\nu_{#1} \to \bar\nu_{#2}}\xspace}
\newcommand{\numubartonuebar}{\nuxtonuxbar{\mu}{e}}

\newcommand{\ptoknubar}{\ensuremath{p \rightarrow K^+ \overline{\nu}}\xspace}

\newcommand{\nnbar}{\ensuremath{n-\bar{n}}\xspace}
\newcommand{\ntoek}{\ensuremath{n\rightarrow e^{-}K^{+}}\xspace}

\newcommand{\snowglobes}{SNOwGLoBES\xspace}

\def\argon40{$^{40}$Ar}  %%%%%       <--- \Ar40 doesn't work; don't know why
\def\Ar39{$^{39}$Ar}
\def\Cl40{$^{40}$Cl}
\def\K40{$^{40}$K}
\def\B8{$^{8}$B}
\newcommand{\lsim}{{\;\raise0.3ex\hbox{$<$\kern-0.75em\raise-1.1ex\hbox{$\sim$}}\;}}
\newcommand{\gsim}{{\;\raise0.3ex\hbox{$>$\kern-0.75em\raise-1.1ex\hbox{$\sim$}}\;}}
\newcommand{\beq}{\begin{equation}}
\newcommand{\eeq}{\end{equation}}
\newcommand{\bea}{\begin{eqnarray}}
\newcommand{\eea}{\end{eqnarray}}

\mathchardef\minus="002D

\newcommand{\snowmasstitle}{
\rule[-0.2in]{\hsize}{0.01in}\\\rule{\hsize}{0.01in}\\
\vskip 0.1in Submitted to the  Proceedings of the US Community Study\\ 
on the Future of Particle Physics (Snowmass 2021)\\ 
\rule{\hsize}{0.01in}\\\rule[+0.2in]{\hsize}{0.01in}
}

%% file: common/init.tex
% This should be \input first thing after \begin{document}

\pagestyle{titlepage}

%\date{14 March, 2022}

\title{\scshape\Large Snowmass Neutrino Frontier:\\
DUNE Physics Summary\\
%\vspace{5mm}
\normalsize Executive Summary of DUNE Physics Program
\vskip -10pt
\snowmasstitle
}

%\vspace{10mm}

%\renewcommand\Authfont{\scshape\small}
%\renewcommand\Affilfont{\itshape\footnotesize}

\newcommand{\Abilene}{Abilene Christian University, Abilene, TX 79601, USA}
\newcommand{\Albanysuny}{University of Albany, SUNY, Albany, NY 12222, USA}
\newcommand{\Amsterdam}{University of Amsterdam, NL-1098 XG Amsterdam, The Netherlands}
\newcommand{\Antalya}{Antalya Bilim University, 07190 D{\"o}{\c{s}}emealt{\i}/Antalya, Turkey}
\newcommand{\Antananarivo}{University of Antananarivo, Antananarivo 101, Madagascar}
\newcommand{\AntonioNarino}{Universidad Antonio Nari{\~n}o, Bogot{\'a}, Colombia}
\newcommand{\Argonne}{Argonne National Laboratory, Argonne, IL 60439, USA}
\newcommand{\Arizona}{University of Arizona, Tucson, AZ 85721, USA}
\newcommand{\Asuncion}{Universidad Nacional de Asunci{\'o}n, San Lorenzo, Paraguay}
\newcommand{\Athens}{University of Athens, Zografou GR 157 84, Greece}
\newcommand{\Atlantico}{Universidad del Atl{\'a}ntico, Barranquilla, Atl{\'a}ntico, Colombia}
\newcommand{\Augustana}{Augustana University, Sioux Falls, SD 57197, USA}
\newcommand{\Basel}{University of Basel, CH-4056 Basel, Switzerland}
\newcommand{\Bern}{University of Bern, CH-3012 Bern, Switzerland}
\newcommand{\Beykent}{Beykent University, Istanbul, Turkey}
\newcommand{\Birmingham}{University of Birmingham, Birmingham B15 2TT, United Kingdom}
\newcommand{\BolognaUniversity}{Universit{\`a} del Bologna, 40127 Bologna, Italy}
\newcommand{\Boston}{Boston University, Boston, MA 02215, USA}
\newcommand{\Bristol}{University of Bristol, Bristol BS8 1TL, United Kingdom}
\newcommand{\Brookhaven}{Brookhaven National Laboratory, Upton, NY 11973, USA}
\newcommand{\Bucharest}{University of Bucharest, Bucharest, Romania}
\newcommand{\CBPF}{Centro Brasileiro de Pesquisas F\'isicas, Rio de Janeiro, RJ 22290-180, Brazil}
\newcommand{\CEASaclay}{IRFU, CEA, Universit{\'e} Paris-Saclay, F-91191 Gif-sur-Yvette, France}
\newcommand{\CERN}{CERN, The European Organization for Nuclear Research, 1211 Meyrin, Switzerland}
\newcommand{\CIEMAT}{CIEMAT, Centro de Investigaciones Energ{\'e}ticas, Medioambientales y Tecnol{\'o}gicas, E-28040 Madrid, Spain}
\newcommand{\CUSB}{Central University of South Bihar, Gaya, 824236, India }
\newcommand{\CalBerkeley}{University of California Berkeley, Berkeley, CA 94720, USA}
\newcommand{\CalDavis}{University of California Davis, Davis, CA 95616, USA}
\newcommand{\CalIrvine}{University of California Irvine, Irvine, CA 92697, USA}
\newcommand{\CalLosangeles}{University of California Los Angeles, Los Angeles, CA 90095, USA}
\newcommand{\CalRiverside}{University of California Riverside, Riverside CA 92521, USA}
\newcommand{\CalSantabarbara}{University of California Santa Barbara, Santa Barbara, California 93106 USA}
\newcommand{\Caltech}{California Institute of Technology, Pasadena, CA 91125, USA}
\newcommand{\Cambridge}{University of Cambridge, Cambridge CB3 0HE, United Kingdom}
\newcommand{\Campinas}{Universidade Estadual de Campinas, Campinas - SP, 13083-970, Brazil}
\newcommand{\CataniaUniversitadi}{Universit{\`a} di Catania, 2 - 95131 Catania, Italy}
\newcommand{\Catolica}{Universidad Cat{\'o}lica del Norte, Antofagasta, Chile}
\newcommand{\Charles}{Institute of Particle and Nuclear Physics of the Faculty of Mathematics and Physics of the Charles University, 180 00 Prague 8, Czech Republic }
\newcommand{\Chicago}{University of Chicago, Chicago, IL 60637, USA}
\newcommand{\ChungAng}{Chung-Ang University, Seoul 06974, South Korea}
\newcommand{\Cincinnati}{University of Cincinnati, Cincinnati, OH 45221, USA}
\newcommand{\Cinvestav}{Centro de Investigaci{\'o}n y de Estudios Avanzados del Instituto Polit{\'e}cnico Nacional (Cinvestav), Mexico City, Mexico}
\newcommand{\Colima}{Universidad de Colima, Colima, Mexico}
\newcommand{\ColoradoBoulder}{University of Colorado Boulder, Boulder, CO 80309, USA}
\newcommand{\ColoradoState}{Colorado State University, Fort Collins, CO 80523, USA}
\newcommand{\Columbia}{Columbia University, New York, NY 10027, USA}
\newcommand{\Cti}{Centro de Tecnologia da Informacao Renato Archer, Amarais - Campinas, SP - CEP 13069-901}
\newcommand{\CzechAcademyofSciences}{Institute of Physics, Czech Academy of Sciences, 182 00 Prague 8, Czech Republic}
\newcommand{\CzechTechnical}{Czech Technical University, 115 19 Prague 1, Czech Republic}
\newcommand{\DakotaState}{Dakota State University, Madison, SD 57042, USA}
\newcommand{\Dallas}{University of Dallas, Irving, TX 75062-4736, USA}
\newcommand{\DannecyleVieux}{Laboratoire d{\textquoteright}Annecy de Physique des Particules, Univ. Grenoble Alpes, Univ. Savoie Mont Blanc, CNRS, LAPP-IN2P3, 74000 Annecy, France}
\newcommand{\Daresbury}{Daresbury Laboratory, Cheshire WA4 4AD, United Kingdom}
\newcommand{\Drexel}{Drexel University, Philadelphia, PA 19104, USA}
\newcommand{\Duke}{Duke University, Durham, NC 27708, USA}
\newcommand{\Durham}{Durham University, Durham DH1 3LE, United Kingdom}
\newcommand{\EIA}{Universidad EIA, Envigado, Antioquia, Colombia}
\newcommand{\ETH}{ETH Zurich, Zurich, Switzerland}
\newcommand{\Edinburgh}{University of Edinburgh, Edinburgh EH8 9YL, United Kingdom}
\newcommand{\Eotvos}{E{\"o}tv{\"o}s Lor{\'a}nd University, 1053 Budapest, Hungary}
\newcommand{\FCULport}{Faculdade de Ci{\^e}ncias da Universidade de Lisboa - FCUL, 1749-016 Lisboa, Portugal}
\newcommand{\FederaldeAlfenas}{Universidade Federal de Alfenas, Po{\c{c}}os de Caldas - MG, 37715-400, Brazil}
\newcommand{\FederaldeGoias}{Universidade Federal de Goias, Goiania, GO 74690-900, Brazil}
\newcommand{\FederaldeSaoCarlos}{Universidade Federal de S{\~a}o Carlos, Araras - SP, 13604-900, Brazil}
\newcommand{\FederaldoABC}{Universidade Federal do ABC, Santo Andr{\'e} - SP, 09210-580, Brazil}
\newcommand{\FederaldoRio}{Universidade Federal do Rio de Janeiro,  Rio de Janeiro - RJ, 21941-901, Brazil}
\newcommand{\Fermi}{Fermi National Accelerator Laboratory, Batavia, IL 60510, USA}
\newcommand{\Ferrarauniv}{University of Ferrara, Ferrara, Italy}
\newcommand{\Florida}{University of Florida, Gainesville, FL 32611-8440, USA}
\newcommand{\Fluminense}{Fluminense Federal University, 9 Icara{\'\i} Niter{\'o}i - RJ, 24220-900, Brazil }
\newcommand{\Genova}{Universit{\`a} degli Studi di Genova, Genova, Italy}
\newcommand{\Georgian}{Georgian Technical University, Tbilisi, Georgia}
\newcommand{\GranSasso}{Gran Sasso Science Institute, L'Aquila, Italy}
\newcommand{\GranSassoLab}{Laboratori Nazionali del Gran Sasso, L'Aquila AQ, Italy}
\newcommand{\Granada}{University of Granada {\&} CAFPE, 18002 Granada, Spain}
\newcommand{\Grenoble}{University Grenoble Alpes, CNRS, Grenoble INP, LPSC-IN2P3, 38000 Grenoble, France}
\newcommand{\Guanajuato}{Universidad de Guanajuato, Guanajuato, C.P. 37000, Mexico}
\newcommand{\Harish}{Harish-Chandra Research Institute, Jhunsi, Allahabad 211 019, India}
\newcommand{\Harvard}{Harvard University, Cambridge, MA 02138, USA}
\newcommand{\Hawaii}{University of Hawaii, Honolulu, HI 96822, USA}
\newcommand{\Houston}{University of Houston, Houston, TX 77204, USA}
\newcommand{\Hyderabad}{University of  Hyderabad, Gachibowli, Hyderabad - 500 046, India}
\newcommand{\IFAE}{Institut de F{\'\i}sica d{\textquoteright}Altes Energies (IFAE){\textemdash}Barcelona Institute of Science and Technology (BIST), Barcelona, Spain}
\newcommand{\IFIC}{Instituto de F{\'\i}sica Corpuscular, CSIC and Universitat de Val{\`e}ncia, 46980 Paterna, Valencia, Spain}
\newcommand{\IGFAE}{Instituto Galego de Fisica de Altas Enerxias, A Coru{\~n}a, Spain}
\newcommand{\INFNBologna}{Istituto Nazionale di Fisica Nucleare Sezione di Bologna, 40127 Bologna BO, Italy}
\newcommand{\INFNCatania}{Istituto Nazionale di Fisica Nucleare Sezione di Catania, I-95123 Catania, Italy}
\newcommand{\INFNFerrara}{Istituto Nazionale di Fisica Nucleare Sezione di Ferrara, I-44122 Ferrara, Italy}
\newcommand{\INFNGenova}{Istituto Nazionale di Fisica Nucleare Sezione di Genova, 16146 Genova GE, Italy}
\newcommand{\INFNLecce}{Istituto Nazionale di Fisica Nucleare Sezione di Lecce, 73100 - Lecce, Italy}
\newcommand{\INFNMilanBicocca}{Istituto Nazionale di Fisica Nucleare Sezione di Milano Bicocca, 3 - I-20126 Milano, Italy}
\newcommand{\INFNMilano}{Istituto Nazionale di Fisica Nucleare Sezione di Milano, 20133 Milano, Italy}
\newcommand{\INFNNapoli}{Istituto Nazionale di Fisica Nucleare Sezione di Napoli, I-80126 Napoli, Italy}
\newcommand{\INFNPadova}{Istituto Nazionale di Fisica Nucleare Sezione di Padova, 35131 Padova, Italy}
\newcommand{\INFNPavia}{Istituto Nazionale di Fisica Nucleare Sezione di Pavia,  I-27100 Pavia, Italy}
\newcommand{\INFNRoma}{Istituto Nazionale di Fisica Nucleare Sezione di Roma, 00185 Roma RM}
\newcommand{\INFNSud}{Istituto Nazionale di Fisica Nucleare Laboratori Nazionali del Sud, 95123 Catania, Italy}
\newcommand{\INR}{Institute for Nuclear Research of the Russian Academy of Sciences, Moscow 117312, Russia}
\newcommand{\IPLyon}{Institut de Physique des 2 Infinis de Lyon, 69622 Villeurbanne, France}
\newcommand{\IPM}{Institute for Research in Fundamental Sciences, Tehran, Iran}
\newcommand{\ISTlisboa}{Instituto Superior T{\'e}cnico - IST, Universidade de Lisboa, Portugal}
\newcommand{\Idaho}{Idaho State University, Pocatello, ID 83209, USA}
\newcommand{\Illinoisinstitute}{Illinois Institute of Technology, Chicago, IL 60616, USA}
\newcommand{\Imperial}{Imperial College of Science Technology and Medicine, London SW7 2BZ, United Kingdom}
\newcommand{\IndGuwahati}{Indian Institute of Technology Guwahati, Guwahati, 781 039, India}
\newcommand{\IndHyderabad}{Indian Institute of Technology Hyderabad, Hyderabad, 502285, India}
\newcommand{\Indiana}{Indiana University, Bloomington, IN 47405, USA}
\newcommand{\Ingenieria}{Universidad Nacional de Ingenier{\'\i}a, Lima 25, Per{\'u}}
\newcommand{\Insubria }{University of Insubria, Via Ravasi, 2, 21100 Varese VA, Italy}
\newcommand{\Iowa}{University of Iowa, Iowa City, IA 52242, USA}
\newcommand{\IowaState}{Iowa State University, Ames, Iowa 50011, USA}
\newcommand{\Iwate}{Iwate University, Morioka, Iwate 020-8551, Japan}
\newcommand{\JINR}{Joint Institute for Nuclear Research, Dzhelepov Laboratory of Nuclear Problems 6 Joliot-Curie, Dubna, Moscow Region, 141980 RU }
\newcommand{\Jammu}{University of Jammu, Jammu-180006, India}
\newcommand{\Jawaharlal}{Jawaharlal Nehru University, New Delhi 110067, India}
\newcommand{\Jeonbuk}{Jeonbuk National University, Jeonrabuk-do 54896, South Korea}
\newcommand{\Jyvaskyla}{University of Jyvaskyla, FI-40014, Finland}
\newcommand{\KEK}{High Energy Accelerator Research Organization (KEK), Ibaraki, 305-0801, Japan}
\newcommand{\KISTI}{Korea Institute of Science and Technology Information, Daejeon, 34141, South Korea}
\newcommand{\KL}{K L University, Vaddeswaram, Andhra Pradesh 522502, India}
\newcommand{\Kansasstate}{Kansas State University, Manhattan, KS 66506, USA}
\newcommand{\Kavli}{Kavli Institute for the Physics and Mathematics of the Universe, Kashiwa, Chiba 277-8583, Japan}
\newcommand{\Kure}{National Institute of Technology, Kure College, Hiroshima, 737-8506, Japan}
\newcommand{\Kyiv}{Taras Shevchenko National University of Kyiv, 01601 Kyiv, Ukraine}
\newcommand{\LIP}{Laborat{\'o}rio de Instrumenta{\c{c}}{\~a}o e F{\'\i}sica Experimental de Part{\'\i}culas, 1649-003 Lisboa and 3004-516 Coimbra, Portugal}
\newcommand{\Lancaster}{Lancaster University, Lancaster LA1 4YB, United Kingdom}
\newcommand{\LawrenceBerkeley}{Lawrence Berkeley National Laboratory, Berkeley, CA 94720, USA}
\newcommand{\Liverpool}{University of Liverpool, L69 7ZE, Liverpool, United Kingdom}
\newcommand{\LosAlmos}{Los Alamos National Laboratory, Los Alamos, NM 87545, USA}
\newcommand{\Louisanastate}{Louisiana State University, Baton Rouge, LA 70803, USA}
\newcommand{\Lucknow}{University of Lucknow, Uttar Pradesh 226007, India}
\newcommand{\Madrid}{Madrid Autonoma University and IFT UAM/CSIC, 28049 Madrid, Spain}
\newcommand{\Mainz}{Johannes Gutenberg-Universit{\"a}t Mainz, 55122 Mainz, Germany}
\newcommand{\Manchester}{University of Manchester, Manchester M13 9PL, United Kingdom}
\newcommand{\Massinsttech}{Massachusetts Institute of Technology, Cambridge, MA 02139, USA}
\newcommand{\Maxplanck}{Max-Planck-Institut, Munich, 80805, Germany}
\newcommand{\Medellin}{University of Medell{\'\i}n, Medell{\'\i}n, 050026 Colombia }
\newcommand{\Michigan}{University of Michigan, Ann Arbor, MI 48109, USA}
\newcommand{\Michiganstate}{Michigan State University, East Lansing, MI 48824, USA}
\newcommand{\MilanoBicocca}{Universit{\`a} del Milano-Bicocca, 20126 Milano, Italy}
\newcommand{\MilanoUniv}{Universit{\`a} degli Studi di Milano, I-20133 Milano, Italy}
\newcommand{\Minnduluth}{University of Minnesota Duluth, Duluth, MN 55812, USA}
\newcommand{\Minntwin}{University of Minnesota Twin Cities, Minneapolis, MN 55455, USA}
\newcommand{\Mississippi}{University of Mississippi, University, MS 38677 USA}
\newcommand{\Newmexico}{University of New Mexico, Albuquerque, NM 87131, USA}
\newcommand{\Niewodniczanski}{H. Niewodnicza{\'n}ski Institute of Nuclear Physics, Polish Academy of Sciences, Cracow, Poland}
\newcommand{\Nikhef}{Nikhef National Institute of Subatomic Physics, 1098 XG Amsterdam, Netherlands}
\newcommand{\Northdakota}{University of North Dakota, Grand Forks, ND 58202-8357, USA}
\newcommand{\Northernillinois}{Northern Illinois University, DeKalb, IL 60115, USA}
\newcommand{\Northwestern}{Northwestern University, Evanston, Il 60208, USA}
\newcommand{\NotreDame}{University of Notre Dame, Notre Dame, IN 46556, USA}
\newcommand{\Occidental}{Occidental College, Los Angeles, CA  90041}
\newcommand{\Ohiostate}{Ohio State University, Columbus, OH 43210, USA}
\newcommand{\OregonState}{Oregon State University, Corvallis, OR 97331, USA}
\newcommand{\Oxford}{University of Oxford, Oxford, OX1 3RH, United Kingdom}
\newcommand{\PacificNorthwest}{Pacific Northwest National Laboratory, Richland, WA 99352, USA}
\newcommand{\Padova}{Universt{\`a} degli Studi di Padova, I-35131 Padova, Italy}
\newcommand{\Panjab}{Panjab University, Chandigarh, 160014 U.T., India}
\newcommand{\Parissaclay}{Universit{\'e} Paris-Saclay, CNRS/IN2P3, IJCLab, 91405 Orsay, France}
\newcommand{\Parisuniversite}{Universit{\'e} de Paris, CNRS, Astroparticule et Cosmologie, F-75006, Paris, France}
\newcommand{\Parma}{University of Parma,  43121 Parma PR, Italy}
\newcommand{\Pavia}{Universit{\`a} degli Studi di Pavia, 27100 Pavia PV, Italy}
\newcommand{\Penn}{University of Pennsylvania, Philadelphia, PA 19104, USA}
\newcommand{\PennState}{Pennsylvania State University, University Park, PA 16802, USA}
\newcommand{\PhysicalResearchLaboratory}{Physical Research Laboratory, Ahmedabad 380 009, India}
\newcommand{\Pisa}{Universit{\`a} di Pisa, I-56127 Pisa, Italy}
\newcommand{\Pitt}{University of Pittsburgh, Pittsburgh, PA 15260, USA}
\newcommand{\Pontificia}{Pontificia Universidad Cat{\'o}lica del Per{\'u}, Lima, Per{\'u}}
\newcommand{\PuertoRico}{University of Puerto Rico, Mayaguez 00681, Puerto Rico, USA}
\newcommand{\Punjab}{Punjab Agricultural University, Ludhiana 141004, India}
\newcommand{\QMUL}{Queen Mary University of London, London E1 4NS, United Kingdom }
\newcommand{\Radboud}{Radboud University, NL-6525 AJ Nijmegen, Netherlands}
\newcommand{\Rochester}{University of Rochester, Rochester, NY 14627, USA}
\newcommand{\Royalholloway}{Royal Holloway College London, TW20 0EX, United Kingdom}
\newcommand{\Rutgers}{Rutgers University, Piscataway, NJ, 08854, USA}
\newcommand{\Rutherford}{STFC Rutherford Appleton Laboratory, Didcot OX11 0QX, United Kingdom}
\newcommand{\SLAC}{SLAC National Accelerator Laboratory, Menlo Park, CA 94025, USA}
\newcommand{\SURF}{Sanford Underground Research Facility, Lead, SD, 57754, USA}
\newcommand{\Salento}{Universit{\`a} del Salento, 73100 Lecce, Italy}
\newcommand{\Sanjosestate}{San Jose State University, San Jos{\'e}, CA 95192-0106, USA}
\newcommand{\Sapienza}{Sapienza University of Rome, 00185 Roma RM, Italy}
\newcommand{\SergioArboleda}{Universidad Sergio Arboleda, 11022 Bogot{\'a}, Colombia}
\newcommand{\Sheffield}{University of Sheffield, Sheffield S3 7RH, United Kingdom}
\newcommand{\SouthDakotaSchool}{South Dakota School of Mines and Technology, Rapid City, SD 57701, USA}
\newcommand{\SouthDakotaState}{South Dakota State University, Brookings, SD 57007, USA}
\newcommand{\Southcarolina}{University of South Carolina, Columbia, SC 29208, USA}
\newcommand{\SouthernMethodist}{Southern Methodist University, Dallas, TX 75275, USA}
\newcommand{\StonyBrook}{Stony Brook University, SUNY, Stony Brook, NY 11794, USA}
\newcommand{\Sunyatsen}{Sun Yat-Sen University, Guangzhou, 510275}
\newcommand{\Sussex}{University of Sussex, Brighton, BN1 9RH, United Kingdom}
\newcommand{\Syracuse}{Syracuse University, Syracuse, NY 13244, USA}
\newcommand{\Tecnologica }{Universidade Tecnol{\'o}gica Federal do Paran{\'a}, Curitiba, Brazil}
\newcommand{\TexasAMcollege}{Texas A{\&}M University, College Station, Texas 77840}
\newcommand{\TexasAMcorpuscristi}{Texas A{\&}M University - Corpus Christi, Corpus Christi, TX 78412, USA}
\newcommand{\TexasArlington}{University of Texas at Arlington, Arlington, TX 76019, USA}
\newcommand{\Texasaustin}{University of Texas at Austin, Austin, TX 78712, USA}
\newcommand{\Toronto}{University of Toronto, Toronto, Ontario M5S 1A1, Canada}
\newcommand{\Tufts}{Tufts University, Medford, MA 02155, USA}
\newcommand{\UNIST}{Ulsan National Institute of Science and Technology, Ulsan 689-798, South Korea}
\newcommand{\Unifesp}{Universidade Federal de S{\~a}o Paulo, 09913-030, S{\~a}o Paulo, Brazil}
\newcommand{\UniversityCollegeLondon}{University College London, London, WC1E 6BT, United Kingdom}
\newcommand{\ValleyCity}{Valley City State University, Valley City, ND 58072, USA}
\newcommand{\VariableEnergy}{Variable Energy Cyclotron Centre, 700 064 West Bengal, India}
\newcommand{\VirginiaTech}{Virginia Tech, Blacksburg, VA 24060, USA}
\newcommand{\Warsaw}{University of Warsaw, 02-093 Warsaw, Poland}
\newcommand{\Warwick}{University of Warwick, Coventry CV4 7AL, United Kingdom}
\newcommand{\Wellesley}{Wellesley College, Wellesley, MA 02481, USA}
\newcommand{\Wichita}{Wichita State University, Wichita, KS 67260, USA}
\newcommand{\WilliamMary}{William and Mary, Williamsburg, VA 23187, USA}
\newcommand{\Wisconsin}{University of Wisconsin Madison, Madison, WI 53706, USA}
\newcommand{\Yale}{Yale University, New Haven, CT 06520, USA}
\newcommand{\Yerevan}{Yerevan Institute for Theoretical Physics and Modeling, Yerevan 0036, Armenia}
\newcommand{\York}{York University, Toronto M3J 1P3, Canada}
\newcommand{\napoli}{Universit{\`a} degli Studi di Napoli Federico II , 80138 Napoli NA, Italy}
%----------------------------------------------------
% So that institutions appear in alphabetical order:
\affiliation{\Abilene}
\affiliation{\Albanysuny}
\affiliation{\Amsterdam}
\affiliation{\Antalya}
\affiliation{\Antananarivo}
\affiliation{\AntonioNarino}
\affiliation{\Argonne}
\affiliation{\Arizona}
\affiliation{\Asuncion}
\affiliation{\Athens}
\affiliation{\Atlantico}
\affiliation{\Augustana}
\affiliation{\Basel}
\affiliation{\Bern}
\affiliation{\Beykent}
\affiliation{\Birmingham}
\affiliation{\BolognaUniversity}
\affiliation{\Boston}
\affiliation{\Bristol}
\affiliation{\Brookhaven}
\affiliation{\Bucharest}
\affiliation{\CBPF}
\affiliation{\CEASaclay}
\affiliation{\CERN}
\affiliation{\CIEMAT}
\affiliation{\CUSB}
\affiliation{\CalBerkeley}
\affiliation{\CalDavis}
\affiliation{\CalIrvine}
\affiliation{\CalLosangeles}
\affiliation{\CalRiverside}
\affiliation{\CalSantabarbara}
\affiliation{\Caltech}
\affiliation{\Cambridge}
\affiliation{\Campinas}
\affiliation{\CataniaUniversitadi}
\affiliation{\Catolica}
\affiliation{\Charles}
\affiliation{\Chicago}
\affiliation{\ChungAng}
\affiliation{\Cincinnati}
\affiliation{\Cinvestav}
\affiliation{\Colima}
\affiliation{\ColoradoBoulder}
\affiliation{\ColoradoState}
\affiliation{\Columbia}
\affiliation{\Cti}
\affiliation{\CzechAcademyofSciences}
\affiliation{\CzechTechnical}
\affiliation{\DakotaState}
\affiliation{\Dallas}
\affiliation{\DannecyleVieux}
\affiliation{\Daresbury}
\affiliation{\Drexel}
\affiliation{\Duke}
\affiliation{\Durham}
\affiliation{\EIA}
\affiliation{\ETH}
\affiliation{\Edinburgh}
\affiliation{\Eotvos}
\affiliation{\FCULport}
\affiliation{\FederaldeAlfenas}
\affiliation{\FederaldeGoias}
\affiliation{\FederaldeSaoCarlos}
\affiliation{\FederaldoABC}
\affiliation{\FederaldoRio}
\affiliation{\Fermi}
\affiliation{\Ferrarauniv}
\affiliation{\Florida}
\affiliation{\Fluminense}
\affiliation{\Genova}
\affiliation{\Georgian}
\affiliation{\GranSasso}
\affiliation{\GranSassoLab}
\affiliation{\Granada}
\affiliation{\Grenoble}
\affiliation{\Guanajuato}
\affiliation{\Harish}
\affiliation{\Harvard}
\affiliation{\Hawaii}
\affiliation{\Houston}
\affiliation{\Hyderabad}
\affiliation{\IFAE}
\affiliation{\IFIC}
\affiliation{\IGFAE}
\affiliation{\INFNBologna}
\affiliation{\INFNCatania}
\affiliation{\INFNFerrara}
\affiliation{\INFNGenova}
\affiliation{\INFNLecce}
\affiliation{\INFNMilanBicocca}
\affiliation{\INFNMilano}
\affiliation{\INFNNapoli}
\affiliation{\INFNPadova}
\affiliation{\INFNPavia}
\affiliation{\INFNRoma}
\affiliation{\INFNSud}
\affiliation{\INR}
\affiliation{\IPLyon}
\affiliation{\IPM}
\affiliation{\ISTlisboa}
\affiliation{\Idaho}
\affiliation{\Illinoisinstitute}
\affiliation{\Imperial}
\affiliation{\IndGuwahati}
\affiliation{\IndHyderabad}
\affiliation{\Indiana}
\affiliation{\Ingenieria}
\affiliation{\Insubria }
\affiliation{\Iowa}
\affiliation{\IowaState}
\affiliation{\Iwate}
\affiliation{\JINR}
\affiliation{\Jammu}
\affiliation{\Jawaharlal}
\affiliation{\Jeonbuk}
\affiliation{\Jyvaskyla}
\affiliation{\KEK}
\affiliation{\KISTI}
\affiliation{\KL}
\affiliation{\Kansasstate}
\affiliation{\Kavli}
\affiliation{\Kure}
\affiliation{\Kyiv}
\affiliation{\LIP}
\affiliation{\Lancaster}
\affiliation{\LawrenceBerkeley}
\affiliation{\Liverpool}
\affiliation{\LosAlmos}
\affiliation{\Louisanastate}
\affiliation{\Lucknow}
\affiliation{\Madrid}
\affiliation{\Mainz}
\affiliation{\Manchester}
\affiliation{\Massinsttech}
\affiliation{\Maxplanck}
\affiliation{\Medellin}
\affiliation{\Michigan}
\affiliation{\Michiganstate}
\affiliation{\MilanoBicocca}
\affiliation{\MilanoUniv}
\affiliation{\Minnduluth}
\affiliation{\Minntwin}
\affiliation{\Mississippi}
\affiliation{\napoli}
\affiliation{\Newmexico}
\affiliation{\Niewodniczanski}
\affiliation{\Nikhef}
\affiliation{\Northdakota}
\affiliation{\Northernillinois}
\affiliation{\Northwestern}
\affiliation{\NotreDame}
\affiliation{\Occidental}
\affiliation{\Ohiostate}
\affiliation{\OregonState}
\affiliation{\Oxford}
\affiliation{\PacificNorthwest}
\affiliation{\Padova}
\affiliation{\Panjab}
\affiliation{\Parissaclay}
\affiliation{\Parisuniversite}
\affiliation{\Parma}
\affiliation{\Pavia}
\affiliation{\Penn}
\affiliation{\PennState}
\affiliation{\PhysicalResearchLaboratory}
\affiliation{\Pisa}
\affiliation{\Pitt}
\affiliation{\Pontificia}
\affiliation{\PuertoRico}
\affiliation{\Punjab}
\affiliation{\QMUL}
\affiliation{\Radboud}
\affiliation{\Rochester}
\affiliation{\Royalholloway}
\affiliation{\Rutgers}
\affiliation{\Rutherford}
\affiliation{\SLAC}
\affiliation{\SURF}
\affiliation{\Salento}
\affiliation{\Sanjosestate}
\affiliation{\Sapienza}
\affiliation{\SergioArboleda}
\affiliation{\Sheffield}
\affiliation{\SouthDakotaSchool}
\affiliation{\SouthDakotaState}
\affiliation{\Southcarolina}
\affiliation{\SouthernMethodist}
\affiliation{\StonyBrook}
\affiliation{\Sunyatsen}
\affiliation{\Sussex}
\affiliation{\Syracuse}
\affiliation{\Tecnologica }
\affiliation{\TexasAMcollege}
\affiliation{\TexasAMcorpuscristi}
\affiliation{\TexasArlington}
\affiliation{\Texasaustin}
\affiliation{\Toronto}
\affiliation{\Tufts}
\affiliation{\UNIST}
\affiliation{\Unifesp}
\affiliation{\UniversityCollegeLondon}
\affiliation{\ValleyCity}
\affiliation{\VariableEnergy}
\affiliation{\VirginiaTech}
\affiliation{\Warsaw}
\affiliation{\Warwick}
\affiliation{\Wellesley}
\affiliation{\Wichita}
\affiliation{\WilliamMary}
\affiliation{\Wisconsin}
\affiliation{\Yale}
\affiliation{\Yerevan}
\affiliation{\York}

%----------------------------------------------------
% Authors in alphabetical order
\author{A.~Abed Abud} \affiliation{\Liverpool}\affiliation{\CERN}
\author{B.~Abi} \affiliation{\Oxford}
\author{R.~Acciarri} \affiliation{\Fermi}
\author{M.~A.~Acero} \affiliation{\Atlantico}
\author{M.~R.~Adames} \affiliation{\Tecnologica }
\author{G.~Adamov} \affiliation{\Georgian}
\author{M.~Adamowski} \affiliation{\Fermi}
\author{D.~Adams} \affiliation{\Brookhaven}
\author{M.~Adinolfi} \affiliation{\Bristol}
\author{C.~Adriano} \affiliation{\Campinas}
% ...
%1244 authors
%----------------------------------------------------
\collaboration{The DUNE Collaboration}
\noaffiliation

\maketitle

\renewcommand{\familydefault}{\sfdefault}
\renewcommand{\thepage}{\roman{page}}
\setcounter{page}{0}

\pagestyle{plain} 
\clearpage
%\textsf{\tableofcontents}

\renewcommand{\thepage}{\arabic{page}}
\setcounter{page}{1}

%\pagestyle{fancy}

% Set how header/footers look
%\newcommand{\chaptermark}[1]{%
%\markboth{Chapter \thechapter:\# 1}{}}
%\renewcommand{\chaptermark}[1]{%
%\markboth{Chapter \thechapter:\ #1}{}}
%\fancyhead{}
%\fancyhead[RO,LE]{\textsf{\footnotesize \thepage}}
%\fancyhead[RO]{\textsf{\footnotesize \thepage}}
%\fancyhead[LO,RE]{\textsf{\footnotesize \rightmark}}
%\fancyhead[LO]{\textsf{\footnotesize \rightmark}}

%\fancyfoot{}
%\fancyfoot[RO]{\textsf{\footnotesize Snowmass 2021}}
%\fancyfoot[LO]{\textsf{\footnotesize Deep Underground Neutrino Experiment}}
%\fancypagestyle{plain}{}

%\renewcommand{\headrule}{\vspace{-4mm}\color[gray]{0.5}{\rule{\headwidth}{0.5pt}}}

% Not all main documents have any citations.
% When not built in "final" mode, add in one citation just to let the
% document build.
% If, after substantial editing a main document still lacks any
% citations then it should have its whole bibliography removed.
%\ifdefined\isfinal\nocite{}\else\nocite{CD0}\fi
%\nocite{nothing}

% see also preamble.tex
%\input{common/acronyms}

%% file: sections/exec-summary.tex
\section*{Executive Summary}
\label{sec:summary}

% CM attempt:
% The Deep Underground Neutrino Experiment (DUNE) is a next-generation long-baseline neutrino oscillation experiment. DUNE's primary physics goal is to directly measure neutrino and antineutrino oscillations as a function of energy in a high-power, broadband beam, a unique capability. This will precisely determine all of the parameters governing long-baseline neutrino oscillation in a single experiment and without degeneracies, and test the three-flavor paradigm. The experimental design has been developed by a large, international collaboration of scientists and engineers to use the exquisite imaging capability of massive LArTPC far detector modules, and to control systematic uncertainties with an argon-based near detector.

The Deep Underground Neutrino Experiment (DUNE) is a next-generation long-baseline neutrino oscillation experiment with a
%with a large, international collaboration of scientists and engineers. DUNE's 
primary physics goal of observing neutrino and antineutrino oscillation patterns to precisely measure 
%all 
the parameters governing long-baseline neutrino oscillation in a single experiment, 
and to test the three-flavor paradigm.
%of $\nu_{1}-\nu_{3}$ and $\nu_{2}-\nu_{3}$ mixing. 
%This is achieved by measuring neutrino and antineutrino oscillations, 
DUNE's 
%experimental 
design has been developed by a large, international collaboration of scientists and engineers to have unique capability to measure neutrino oscillation
as a function of energy in a broadband beam, to resolve degeneracy among oscillation parameters, and to control systematic uncertainty using the exquisite imaging capability of massive LArTPC far detector modules and an argon-based near detector.
DUNE's neutrino oscillation measurements will unambiguously resolve the neutrino mass ordering and provide the sensitivity to discover CP violation in neutrinos for a wide range of possible values of $\deltacp$.
%, in a single experiment.
DUNE is also uniquely sensitive to electron neutrinos from a galactic supernova burst, and to a broad range of physics beyond the Standard Model (BSM), including nucleon decays. DUNE is anticipated to begin collecting physics data 
%in the late 2020s 
with Phase~I, an initial experiment configuration consisting of two far detector modules and a minimal suite of near detector components, with a 1.2~MW proton beam.
%, but with significantly reduced capability relative to the full experiment design. 
In Phase I, DUNE will be able to quickly 
and unambiguously 
determine the neutrino mass
ordering, a capability that is unique among existing and planned experiments, observe CP violation with 3$\sigma$ significance if $\deltacp = -\pi/2$, measure other oscillation parameters including $\Delta m^{2}_{32}$ with world-leading precision, detect neutrinos from a core-collapse supernova, and search for BSM physics. To realize its extensive, world-leading physics potential requires the full scope of DUNE be completed in Phase II. Phase II will facilitate 5$\sigma$ CP violation discovery potential, 7-16 degree resolution on \deltacp, independent measurement of \sinstt{13} with precision comparable to that of reactors, significant sensitivity to the $\theta_{23}$ octant, and a broad physics program with unique and world-leading sensitivity to physics beyond the three neutrino paradigm as well as additional BSM physics and astrophysics. The three Phase II upgrades are all necessary to achieve DUNE's physics goals: (1) addition of far detector modules three and four for a total FD fiducial mass of at least 40 kt, (2) upgrade of the proton beam power from 1.2 MW to 2.4 MW, and (3) replacement of the near detector's temporary muon spectrometer with a magnetized, high-pressure gaseous argon TPC and calorimeter. 
%While the details of the staging scenario have some impact on early sensitivity milestones, 
The precision physics program -- including low-energy and BSM physics -- is not greatly impacted by the details of the staging plan as long as these upgrades are aggressively pursued.
It is critical for the long-term success of this program that the community and P5 re-affirm that the completion of the full DUNE design is the highest priority in the next decade.

%%%%%%%%%%%%%%%%%%%%%%%%%%%%%%%%%%%%%%%%%%%%%%%%%%%%%%%%%%%

%% file: sections/introduction.tex
\section{The Deep Underground Neutrino Experiment}
\label{sect:introduction}

The Deep Underground Neutrino Experiment (DUNE) is an international, next-generation long-baseline neutrino experiment based in the United States, with three primary physics goals:

\begin{enumerate}
\setlength\itemsep{-0.5em}
    \item make precise measurements of the parameters governing $\nu_{1}-\nu_{3}$ and $\nu_{2}-\nu_{3}$ in a single experiment, including the neutrino mass ordering and the CP-violating phase $\deltacp$, and test the three-flavor paradigm,
    \item make astrophysics and particle physics measurements with supernova burst neutrinos and other low-energy neutrinos, and
    \item search for physics beyond the Standard Model.
\end{enumerate}

% Shorter, splashier paragraph about DUNE concept
DUNE is designed to achieve these goals by pushing both the intensity and precision beyond what is achieved by current long-baseline neutrino oscillation experiments. Key features of this design include:

\begin{itemize}
    \item the world's most intense neutrino and antineutrino beams,
    \item a broad neutrino spectrum to measure the shape of the oscillation pattern, 
    %as a function of energy,
    \item large far detectors located deep underground,
    \item liquid argon technology for particle identification and energy measurement, and
    \item an optimized near detector with the same technology as the far detector, to constrain systematics to world-leading precision and enable a broad physics program.
\end{itemize}

DUNE 
%has a broad physics program and 
is poised to make numerous high-impact, world-leading measurements in a broad range of particle and astroparticle physics over the next several decades. Unambiguous determination of the neutrino mass ordering would resolve one of the longstanding questions in neutrino physics, which has significant implications for astrophysics, cosmology, and neutrinoless double beta decay experiments. The observation of CP violation in neutrinos would be an important step in understanding whether leptogenesis is a viable explanation of the baryon asymmetry of the universe. High-precision measurements of mixing parameters will not only complete the three-flavor picture, but also test it for inconsistencies that could point us to physics beyond the three-flavor model. 
Determination of the $\theta_{23}$ octant could point to a previously unknown symmetry. DUNE will detect a large sample of neutrinos from a supernova burst within the Milky Way and possibly beyond, with unique sensitivity to the electron neutrino component, allowing a wide range of astrophysics and particle physics measurements. DUNE can probe a diverse range of phenomenology beyond the Standard Model, including searches for dark matter, sterile neutrino mixing, nonstandard neutrino interactions, CPT violation, new physics enhancing neutrino trident production, and baryon number violating processes.

% progress since the last P5
The 2014 P5 report~\cite{HEPAPSubcommittee:2014bsm} prioritized DUNE's physics program as one of the primary goals for US particle physics. The P5 report describes a phased approach to building the world's most precise neutrino oscillation experiment. It sets a goal of an initial configuration capable of accumulating 120 kt-MW-yrs by the mid 2030s, sufficient to determine the neutrino mass ordering and to discover CP violation at $3\sigma$ if \deltacp is maximal. The report also establishes that the facility should be upgraded to a multi-megawatt beam and at least 40 kt of far detector fiducial mass in order to surpass 600 kt-MW-yrs and enable precision neutrino oscillation measurements. 

\begin{figure}[htbp]
  \centering
  \includegraphics[width=0.9\linewidth]{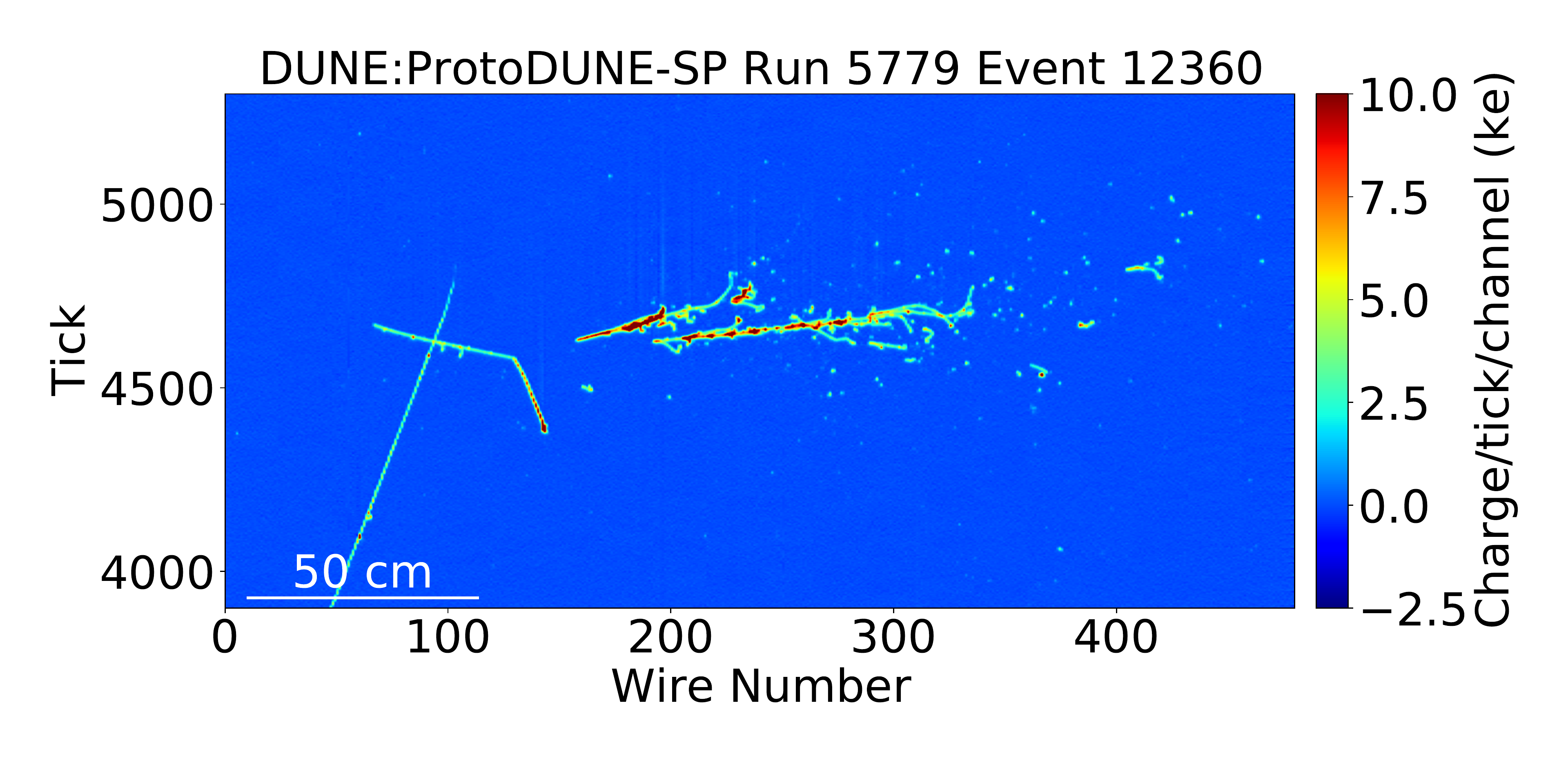}
  \caption{An event display showing a pion charge exchange candidate from ProtoDUNE-SP data. In this event, 3 GeV $\pi^{+}$ from the beamline enters the detector (from left) and undergoes a charge exchange interaction, producing a proton and $\pi^{0}$. The proton can be seen coming to rest, while the $\pi^{0}$ decays into two photons, which induce electromagnetic showers. A cosmic ray muon can be seen overlapping with the incident pion. ProtoDUNE sees a large flux of cosmic rays due to its surface location, but the underground DUNE FD will never see overlapping signals.}
    \label{fig:evdisp}
\end{figure}

Following the recommendations of the P5, DUNE has formed an international collaboration of more than 1000 members from over 200 institutions in more than 30 countries. The US Department of Energy and multiple international partners have invested significant resources in DUNE, including physics research, detector development and prototyping, and infrastructure. Liquid argon TPC technology has been demonstrated in large-scale detectors, with MicroBooNE\cite{MicroBooNE:2021rmx,MicroBooNE:2021cue} and ICARUS\cite{Antonello:2015lea,ICARUS:2021dnd} operating successfully in neutrino beams, and ProtoDUNE running successfully in a hadron test beam at full scale in the drift dimension\cite{DUNE:2020cqd,DUNE:2021hwx}. Figure~\ref{fig:evdisp} is an event display from ProtoDUNE data, which shows the granularity with which individual final-state particles can be detected in DUNE. DUNE has published a technical design report of the far detector~\cite{tdr-vol-1}, and a conceptual design report of the near detector~\cite{DUNE:2021tad}. The collaboration has developed sophisticated physics analyses based on full simulations, including a long-baseline neutrino oscillation analysis with algorithmic event selection and reconstruction, and with realistic flux, cross section, and detector systematic uncertainties included~\cite{tdr-vol-2}. Additional extensive investment has been made in the Long Baseline Neutrino Facility (LBNF), which enables DUNE's physics by providing the neutrino beam and conventional facilities for the detectors. Excavation of the DUNE drifts and caverns at SURF is underway, final designs for the beamline complex and near detector complex have been completed, and the neutrino beamline final design is nearing completion. PIP-II, Fermilab's proton accelerator improvement plan, which is necessary to provide an intense neutrino beam, is well underway, with conventional facilities nearing completion, and is expected to be ready to provide proton beams by 2029.

The full physics program of DUNE requires four far detector modules (40 kt fiducial mass), an initial proton beam intensity of 1.2 MW with an upgrade to 2.4 MW, and a highly capable near detector system. Due to funding constraints, DUNE is expected to begin collecting physics data in the late 2020s, with the neutrino beam available in 2032. The initial detector configuration (Phase I) will include two far detector modules and a limited near detector, as described in Section~\ref{sect:practical}, with a 1.2 MW proton beam. This configuration is sufficient to achieve DUNE's early physics goals, but the full physics program is in jeopardy if the full scope of the design (Phase II) is not realized. %It is critical for the long-term success of this program that the community and P5 re-affirm that the completion of the full LBNF/DUNE design is the highest priority in the next decade.

This document is offered as a summary of DUNE's design considerations and expected physics sensitivity, which have been extensively documented in \cite{Abi:2020evt,DUNE:2020jqi,DUNE:2020zfm,DUNE:2020fgq,DUNE:2021tad,DUNE:2021mtg}. Section~\ref{sect:design} describes each of the fundamental design choices and how it is motivated by DUNE's broad physics goals. Section~\ref{sect:phys} summarizes studies quantifying DUNE's sensitivity to these physics measurements. Section~\ref{sect:practical} discusses practical matters, such as schedules and funding, that will impact DUNE's ability to carry out the physics program described herein and Section~\ref{sect:snowmass} summarizes the collaboration's message to Snowmass.

%%%%%%%%%%%%%%%%%%%%%%%%%%%%%%%%%%%%%%%%%%%%%%%%%%%%%%%%%%%

%% file: sections/expt_design.tex
\section{DUNE Design Considerations}
\label{sect:design}

%%%%%%%%%%%%%%%%%%%%%%%%%%%%%%%%%%%%%%%%%%%%%%%%%%%%%%%%%%%

% Moved from introduction
DUNE will observe neutrinos from the new LBNF~\cite{tdr-vol-1,LBNFBeamline} neutrino beam originating at Fermilab. LBNF will use the high-power %PIP-II~\cite{pip2-2013} 
Fermilab proton accelerator to produce a horn-focused beam of predominately muon neutrinos or antineutrinos with a broad spectrum of neutrino energies that peaks at 2.5~GeV. Measurements will be made with a near detector (ND) located on site at Fermilab and a far detector (FD) located 1285 km from the neutrino production point, at the Sanford Underground Research Facility (SURF), in Lead, South Dakota. The near detector will consist of a suite of detector technologies designed to constrain systematic uncertainties in the oscillation measurements. The far detector will consist of four large liquid argon time projection chamber (LArTPC) modules located at the 4850-ft level of SURF. Combined data from the near and far detectors will allow DUNE to unambiguously determine the neutrino mass ordering, discover CP violation in neutrinos for a wide range of possible values of the CP-violating phase $\deltacp$, and precisely measure $\deltacp$, $\theta_{13}$, $\theta_{23}$, and $\dm{32}$ in a single experiment. These measurements will resolve longstanding questions in neutrino physics, test the consistency of the PMNS framework, and differentiate among models that predict mixing parameters and/or relationships among these parameters. The far detector technology and location deep underground will facilitate study of low-energy neutrinos, including sensitivity to solar neutrinos and neutrinos from a core-collapse supernova, should one occur in our vicinity during the lifetime of the experiment. DUNE will perform searches for baryon number violating processes, such as proton decay and neutron-antineutron annihilation, and for other physics beyond the Standard Model, including both searches for BSM processes observable as deviations from the standard three-flavor neutrino mixing paradigm and direct searches for new particles, including dark matter of cosmic origin.

Design choices in DUNE are largely based on the requirements of the long-baseline neutrino oscillation physics program, though sensitivity to low-energy neutrino interactions such as those from supernova burst neutrinos have also been an important consideration. 
%The goals of the long-baseline neutrino oscillation physics program are to measure the neutrino mass ordering, $\deltacp$, $\theta_{13}$, $\theta_{23}$, and $\Delta m^{2}_{23}$ without ambiguity in a single experiment. Sensitivity to MeV-scale neutrinos such as those from a supernova burst, and to non-accelerator signals such as nucleon decays, are also important considerations. 
In this section, the primary experimental principles of DUNE are summarized, with a brief explanation of how each is driven by physics considerations.

The high-level physics requirements of DUNE are:

\begin{itemize}
    \item {\bf Long baseline:} With a baseline greater than $\sim 1000$~km, the asymmetry in the $\numutonue$ and $\numubartonuebar$ oscillation probabilities due to the matter effect is much larger than that arising from CP violation, regardless of the value of $\deltacp$. This breaks the experimental degeneracy between the neutrino mass ordering and the determination of $\deltacp$\cite{Bass:2013vcg,Abi:2020qib}, allowing DUNE to measure both using the same data set unambiguously, a unique capability among existing and planned neutrino experiments. 
    
    \item {\bf Large far detector mass and high-power beam:} 
    %Neutrinos rarely interact with matter, so to accumulate the necessary statistics for a precision experiment, particularly in the $\nue$ appearance modes, is a challenge. 
    The fiducial mass of the far detector and the intensity of the proton beam linearly impact the number of neutrino interactions. The full FD mass (at least 40~kt fiducial, 20~kt fiducial initially) and full beam power (2.4~MW, 1.2~MW initially) are required to reach the precision goals of the experiment. 
    
    \item {\bf Wide-band neutrino beam:} To definitively establish CP violation, precisely test the three neutrino mixing paradigm, and resolve ambiguities between parameters, DUNE will observe neutrino oscillations as a function of $L/E$ over more than one full oscillation period. At a fixed baseline, this corresponds to measuring neutrinos with a range of energies. The horn-focused neutrino beam line builds on the successful NuMI~\cite{Adamson:2015dkw} beam design by optimizing the focusing parameters to maximize sensitivity to CP violation in DUNE. The beam can be operated in forward (reverse) horn current mode, resulting in a predominately $\numu$ ($\anumu$) flux that is peaked at about 2.5 GeV, near the first oscillation maximum, with significant rate covering more than one oscillation period (0.5-4.0~GeV), less than 5\% contamination from wrong-sign $\anumu$ ($\numu$), and less than 1\% contamination from $\nue$/$\anue$ components.

    \item {\bf Precise neutrino flavor and energy reconstruction:} It is critical to accurately separate $\nu_{\mu}$ and $\nu_{e}$ charged-current interactions in the FD, which requires excellent discrimination between muons and electrons. To measure oscillations as a function of $L/E$ at fixed baseline, the FD must have the ability to reconstruct the neutrino energy precisely, event by event.

    \item {\bf Precise constraints of systematic uncertainties:} Precisely measuring $\deltacp$ requires a few-percent-level measurement of muon-to-electron oscillations, as a function of neutrino energy. Systematic uncertainties due to the neutrino flux prediction, interaction cross sections, and detector response must be constrained at a level better than the statistical precision.

    \item {\bf Deep underground FD location:} The deep underground location is critical for searches for rare non-accelerator signals, due to the high rate of activity from cosmic particles for a detector on the surface. These signals include baryon number violating processes, which require nearly zero background, as well as low-energy neutrinos from a supernova burst. An underground FD also greatly simplifies the reconstruction of accelerator neutrinos.

\end{itemize}

\noindent
These high-level requirements are satisfied by the full DUNE design, which includes 40 kt fiducial mass, deep underground, in a high-power wide-band beam. Below, the physics motivation behind specific detector design choices is described:

\begin{itemize}
    \item {\bf Liquid argon TPC (LArTPC) far detector:} Given the baseline of DUNE, the first oscillation maximum occurs at 2.5 GeV neutrino energy. In this energy regime, many neutrino interaction processes are important, including those with multiple final-state mesons. The LArTPC technology combines tracking and calorimetry, allowing for identification of $\numu$ and $\nue$ CC interactions, as well as good resolution to both the lepton and hadronic energies, and scalability to the required detector mass. The imaging capability of the LArTPC FD also plays an important role in many DUNE BSM searches, and the argon target provides unique sensitivity to electron neutrinos (as opposed to antineutrinos) from supernova bursts. A visualization of a ProtoDUNE pion charge exchange event is shown in Figure~\ref{fig:evdisp}, showing high-resolution imaging of DUNE and the granularity with which individual particles can be measured.
    
    % Newer, longerer paragraph
    \item {\bf Near detector system including LArTPC:} The role of the ND is to measure the neutrino beam prior to the onset of oscillation effects, and to minimize systematic uncertainties in long-baseline oscillation measurements. The observed neutrino energy spectrum depends on a complicated product of flux, cross sections, and detector response, all of which have large \textit{a priori} uncertainties. Because the FD is a LArTPC, the only way to be sensitive to all three of these classes of uncertainty simultaneously is with a LArTPC at the near site, with kinematic acceptance at least as good as the FD. ND-LAr is optimized to meet this requirement, and is critical for all stages of the experiment. In order to cope with the high event rate at the near site, ND-LAr is designed with pixelated readout and optical segmentation. To minimize costs, ND-LAr is not magnetized, and is not large enough to contain forward muons. Instead, a downstream muon spectrometer (the Temporary Muon Spectrometer, TMS, in Phase I) measures the momentum and charge sign of exiting muons. Because neutrino cross sections depend on energy, and the fluxes at the ND and FD are different (mainly due to oscillations), it is important to take ND data in different neutrino fluxes. This is enabled by the PRISM technique, in which ND-LAr and TMS move laterally from the primary proton beam direction. As one moves off axis, the neutrino energy spectrum shifts downward. Data collected with different spectra can be combined to produce a data-driven prediction of the \textit{oscillated} FD spectrum, which is largely independent of interaction and detector modeling. SAND remains on axis, and has a broad and complementary cross section and exotic physics program in addition to monitoring the neutrino beam. To reach the ultimate oscillation physics goals, measurements of neutrino-argon interactions with even greater precision are required. This is fulfilled in Phase II by replacing TMS with ND-GAr, a magnetized high-pressure gaseous argon TPC with a surrounding calorimeter. Its gaseous active target volume gives extremely low thresholds, minimal secondary interactions, and particle-by-particle charge and momentum reconstruction, which extend the reach of the ND. Each of these systems plays a critical role in the experiment. In addition to its critical role in oscillation measurements, the DUNE ND in conjunction with the high-power beam also enables numerous BSM searches, as well as precision measurements of electroweak physics and nuclear structure.

    \item {\bf Calibration: } To achieve the unprecedented level of systematic uncertainty required for DUNE's oscillation physics goals, the position and energy of charge deposits in the far detector must be precisely calibrated at the few percent level. 
    %This is particularly critical because both detector effects and the neutrino interaction model have significant impact on reconstruction of kinematic quantities, so it can be difficult to separately evaluate the impact of each uncertainty. 
    Calibration is challenging in LArTPCs because there are a number of low-level detector properties
    %, including argon purity, argon fluid dynamics, electric field uniformity, diffusion, and electronics effects, 
    that all impact a single high-level quantity such as the energy scale uncertainty. For these reasons, it is important to have as many independent sources of calibration information as possible; DUNE's calibration strategy involves prototype and test beam measurements, analysis of calibration data from physics samples, 
    %such as neutrinos, cosmic rays, and argon isotopes, 
    and detector instrumentation.
    %such as argon purity monitors, a laser calibration system, and a pulsed neutron system.
    
    \item {\bf Photon detection: } The photon detection system is used to observe prompt scintillation light induced by charged particles traversing the LArTPC and thus provides timing information for non-beam physics events. For supernova burst events, where mapping out the time evolution of the neutrino flux is an important physics goal, the photon detection system provides sufficient timing resolution that this measurement is not limited by the detector. The scintillation signal is used in combination with the TPC information to determine the position in the drift direction. The light yield requirement for the photon detection system is set such that the three dimensional vertex resolution in the TPC allows the fiducial volume to be determined at the 1\% level. 
    %The photon detection system can also be used to improve calorimetric energy measurements in the FD.
    
    \item {\bf Data acquisition system:} The trigger and DAQ system is responsible for receiving, processing, and recording data,
    %It receives and buffers data streaming from the TPC and the PDS, carries out data reduction and compression as needed, forms trigger decisions, and stores the corresponding selected detector space-time volumes. 
    and for providing timing and synchronization to the detector electronics and calibration devices.
    The DAQ design requirements are most strongly driven by the goal of detecting supernova burst neutrinos; these include stringent up-time requirements, to increase the chances that the detector is collecting data when one of these rare transient events occurs, as well as data rate and throughput considerations.
    %The activity associated with beam, cosmic rays, and atmospheric neutrinos is localized in space and particularly in time; SNB neutrinos are associated with activity that extends over the entirety of the detector and lasts between 10 and 100 s.
    %Beam, cosmic and nucleon decay events will be triggered using localized regions of visible activity. It is important to accept all possible events with as high an efficiency and as wide a region-of-interest in channel and time space as possible, regardless of event type. An extended low-energy trigger will be used for SNB trigger looking for coincident regions of low-energy deposits, below 10 MeV, across an entire module and in a 10 s period. An extended high-energy trigger will open a readout window of 100 s to capture a full SNB. 
    
\end{itemize}

%This is an example of a figure, Fig.~\ref{fig:disappearance}.

%\begin{figure}[h!]
%    \centering
 %   \includegraphics[height=3.5in,width=0.43\textwidt%h]{graphics/disappearance}
 %   \includegraphics[height=3.7in,width=0.47\textwidt%h]{graphics/ComboLimit99_prelim.pdf}
 %   \caption{Left panel: Comparison of present %exclusion limits from various experiments obtained %through searches for disappearance of muon neutrinos %into sterile species assuming a 3+1 model.  The %Gariazzo et al. region represents a global fit to %neutrino oscillation data~\cite{Gariazzo_2015}. 
 %   Right panel: The combined results of the %disappearance measurements from \minosplus, %\dayabay, and \bugey, compared to the appearance %measurements from \lsnd and \miniboone.}
 %   \label{fig:disappearance}
%\%end{figure}

%%%%%%%%%%%%%%%%%%%%%%%%%%%%%%%%%%%%%%%%%%%%%%%%%%%%%%%%%%%

%% file: sections/phys_sens.tex
\section{DUNE Physics Sensitivity}
\label{sect:phys}

\subsection{Long-Baseline Oscillation Physics}
\begin{figure}[htbp]
  \centering
  \includegraphics[width=0.45\linewidth]{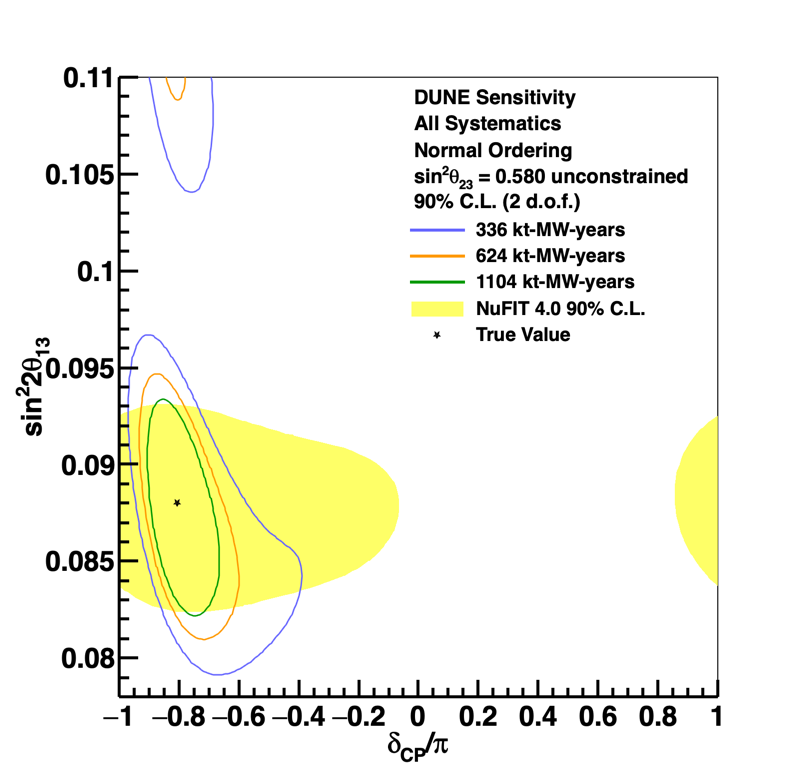}
  \includegraphics[width=0.45\linewidth]{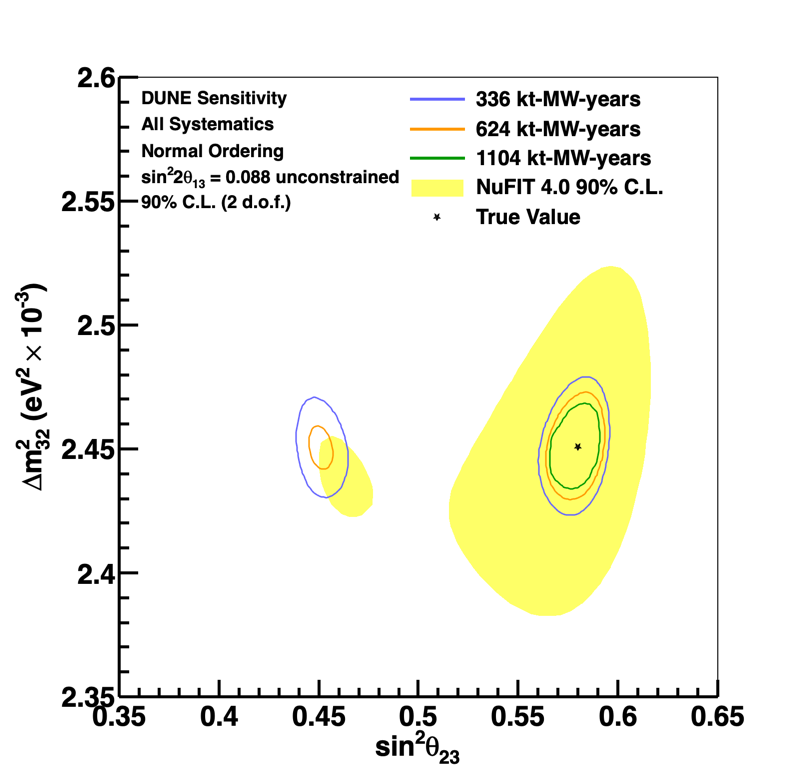}
  \caption{Two-dimensional 90\% C.L. regions in the \sinstt{13}--\deltacp (left) and \sinst{23}--\dm{32} (right)  plane, for three different levels of exposure, with equal running in neutrino and antineutrino mode, with the Phase II near detector. The 90\% C.L. region for the NuFIT global fit is shown in yellow for comparison. The true values of the oscillation parameters are assumed to be the central values of the NuFit global fit and the oscillation parameters governing long-baseline oscillation are unconstrained.}
    \label{fig:sens_precision}
\end{figure}

DUNE has performed analyses of the long-baseline oscillation sensitivity based on full, end-to-end simulation, reconstruction, and event selection of far detector Monte Carlo and parameterized analysis of near detector Monte Carlo\cite{DUNE:2020jqi}. A convolutional neural network has been developed to select $\nue$ and $\numu$ CC events; the selection achieves greater than 85\% efficiency for reconstructed energies between 2-5 GeV\cite{DUNE:2020gpm}. Detailed uncertainties from flux, the neutrino interaction model, and detector effects have been included in the analysis. Sensitivity results are obtained using a sophisticated, custom fitting framework. These studies demonstrate that DUNE will be able to measure \deltacp to high precision, unequivocally determine the neutrino mass ordering, and make precise measurements of the parameters governing long-baseline neutrino oscillation. A demonstration of DUNE's precision measurement capability is shown in Fig.~\ref{fig:sens_precision}; at large exposures DUNE is able to simultaneously determine both $\deltacp$ and \sinst{13}, with the precision on \sinst{13} similar to the current world average dominated by reactor neutrino experiments. The multiple allowed regions present for smaller exposures are eliminated at the highest exposure shown, demonstrating how DUNE will ultimately be able to resolve degeneracy among the parameters.

\begin{figure}[htbp]
  \centering
  \includegraphics[width=0.45\linewidth]{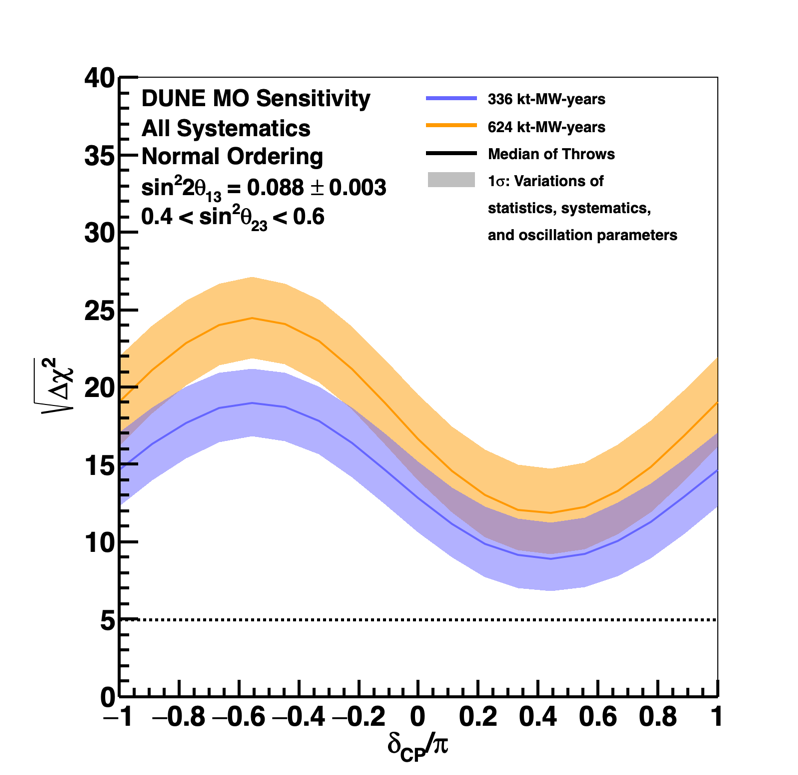}
  \includegraphics[width=0.45\linewidth]{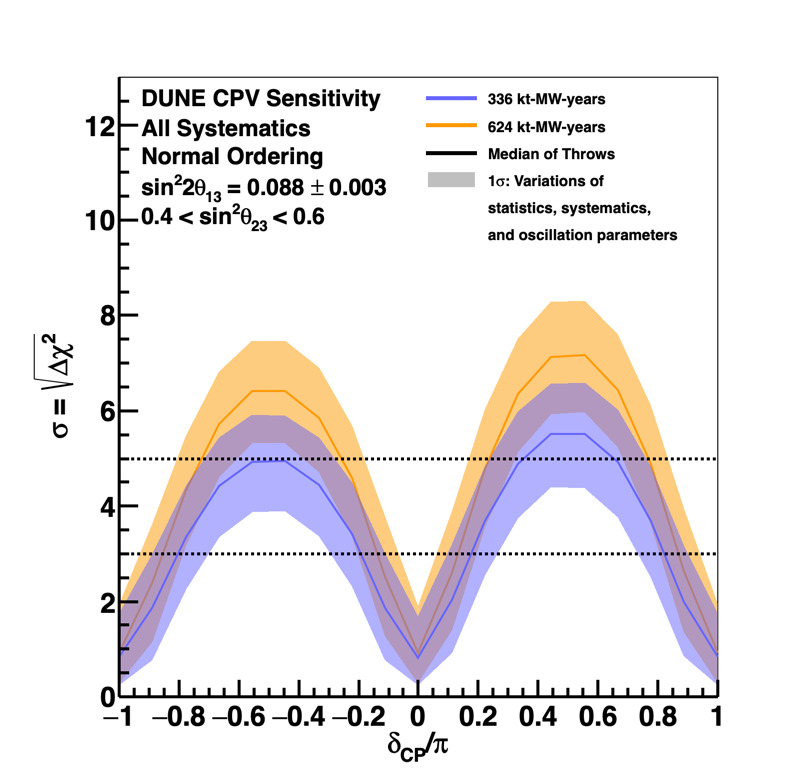}
  \includegraphics[width=0.45\linewidth]{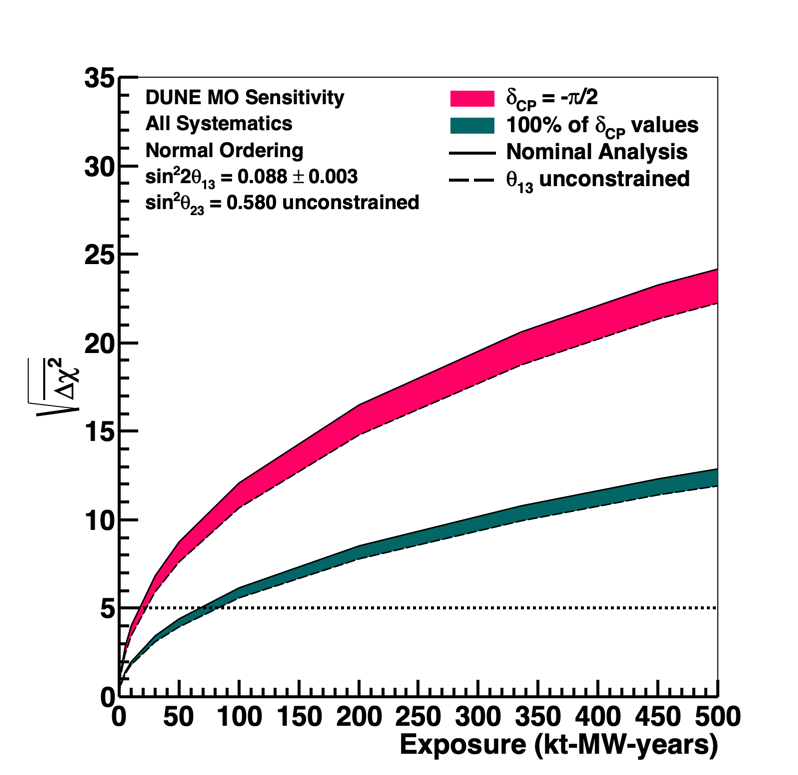}
  \includegraphics[width=0.45\linewidth]{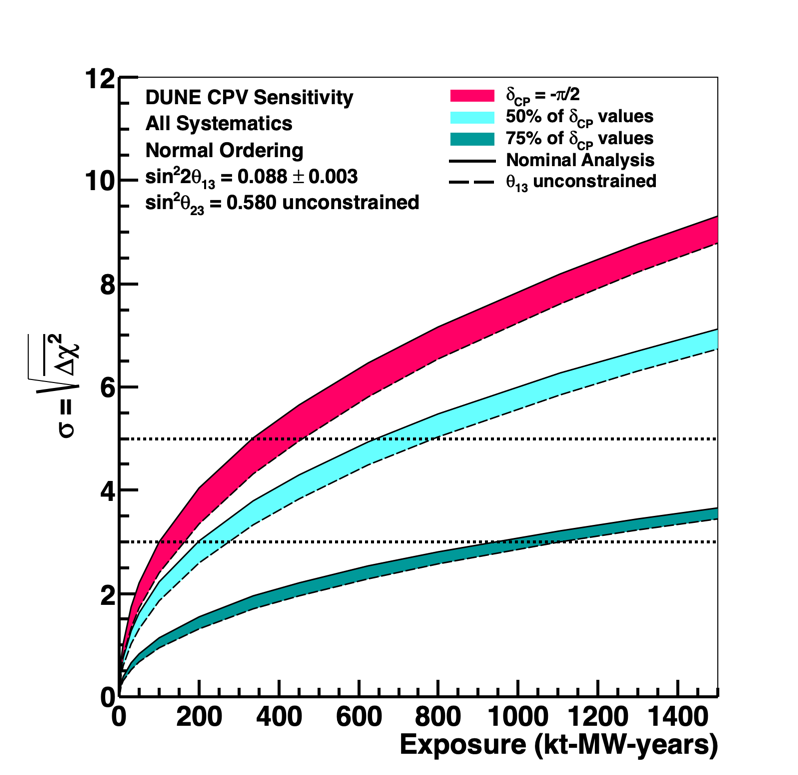}
  \caption{Sensitivity to the neutrino mass ordering (left) and CP violation (right) as a function of the true value of $\deltacp$ for several exposures (top) or as a function of exposure in kt-MW-years (bottom), with the Phase II near detector. For the top plots, the solid line shows the median sensitivity while the width of the band represents 68\% of variations of statistics, systematics, and oscillation parameters. In the bottom plots, only the Asimov sensitivity is considered and the width of the bands represents the difference between the nominal analysis, which uses the external constraint from reactor antineutrino experiments on $\sin^{2}\theta_{13}$, and an analysis without this constraint. Curves are shown for maximal CP violation ($\deltacp=-\pi/2$) and for different fractions of $\deltacp$ values.}
    \label{fig:cpvmhsens}
\end{figure}

DUNE's ultimate oscillation physics goal is precision measurements of all the parameters governing long-baseline neutrino oscillation and sensitivity to deviations from standard three-flavor mixing. Other important physics milestones are unambiguous determination of the neutrino mass ordering and discovery of CP violation for a range of true $\deltacp$ values. Figure~\ref{fig:cpvmhsens} shows the sensitivity to neutrino mass ordering and CP violation as a function of the true value of $\deltacp$ for a given exposure and the sensitivity as a function of exposure for a given $\deltacp$ value or range of values. Precise determination of the parameters governing long-baseline oscillation is necessary to achieve high sensitivity to CP violation over a wide range of possible $\deltacp$ values, so this sensitivity can also be viewed as a proxy for DUNE's precision measurement capability. 

In Phase I, DUNE will be able to establish the neutrino mass ordering at the 5$\sigma$ level for 100\% of \deltacp values with 100 kt-MW-years exposure, and has the ability to make strong statements with significantly shorter exposures depending on the true values of other oscillation parameters. DUNE's capacity to quickly and  unambiguously determine the neutrino mass ordering is unique among existing and planned experiments. DUNE can also observe CP violation with 3$\sigma$ significance with 100 kt-MW-years exposure if $\deltacp = -\pi/2$, which corresponds to maximal CP violation.
Analysis of early DUNE sensitivity is described in \cite{DUNE:2021mtg}.

CP violation can be observed with 5$\sigma$ significance after $\sim$350 kt-MW-years if \deltacp = $-\pi/2$ and after $\sim$700 kt-MW-years for 50\% of \deltacp values. CP violation can be observed with 3$\sigma$ significance for 75\% of \deltacp values after $\sim$1000 kt-MW-years of exposure. For large exposures, \deltacp resolution between five and fifteen degrees is possible, depending on the true value of \deltacp, and the DUNE measurement of \sinstt{13} approaches the precision of reactor experiments. For this reason, ultimately DUNE is able to simultaneously measure all the parameters governing $\nu_{1}-\nu_{3}$ and $\nu_{2}-\nu_{3}$ mixing in a single experiment, without an external \sinstt{13} constraint. This facilitates discrimination among flavor and BSM models that predict oscillation parameters, and a comparison between the DUNE and reactor \sinstt{13}  results, which allows for a test of the unitarity of the PMNS matrix.
%is of interest as a potential signature for physics beyond the standard model. 
DUNE will also have significant sensitivity to the $\theta_{23}$ octant for values of \sinst{23} less than about 0.47 and greater than about 0.55. Conversion of exposures to calendar years depends upon the staging scenario, which is addressed in Section~\ref{sect:practical}, but if upgrades are aggressively pursued, the details of the early staging do not greatly impact this precision physics program.

%Studies demonstrating the importance of near detector constraints and, in particular, robustness against insufficient modeling of neutrino interactions have been performed and analyses demonstrating how DUNE can improve interaction modeling and reduce model dependence have been developed. Details of these studies are provided in the DUNE Near Detector CDR\cite{DUNE:2021tad}. A demonstration of the DUNE PRISM analysis making use of simulated off-axis data is in progress. An active analysis program is responsible for implementing, within the DUNE collaboration, reconstruction, calibration, and broader analysis techniques aimed at further developing detector performance.
DUNE's precision measurement program, including significant sensitivity to CP violation for a wide range of $\deltacp$ values, relies upon control of systematic uncertainties, which is largely achieved using near detector measurements. For this reason, care has been taken to implement a sophisticated treatment of systematic uncertainties in the sensitivity studies presented here, which explicitly include constraints from a joint fit to near and far detector samples on a large number of individual systematic parameters. The analysis includes parameters describing uncertainty in the flux, both from hadron production and focusing, in the neutrino-nuclear interaction model, including effects from initial- and final-state interactions, and from detector effects. Implicit in the sensitivities shown in Figs.~\ref{fig:sens_precision} and \ref{fig:cpvmhsens} is the use of near detector data 
to perform an analysis that is robust against 
%in providing robustness against 
insufficient modeling of neutrino interactions; studies demonstrating the impact of shortcomings in the neutrino interaction model are described in \cite{Abi:2020evt,DUNE:2021tad} and are reflected in some of the studies shown in Section~\ref{sect:practical}. The ability to observe the flux at multiple off-axis angles with the near detector, called DUNE PRISM, is a critical feature of DUNE's design that will reduce DUNE's dependence on the neutrino-nucleus interaction model by facilitating largely data-driven predictions of far detector spectra. Dedicated analyses of near detector samples will contribute to improvement of the neutrino-nucleus interaction model. DUNE's analysis strategy of performing fits to constrain variation in model parameters, incorporating analysis techniques that reduce model dependence, and making improvements to models has been developed to ensure that DUNE is able to control systematic uncertainty sufficiently to reach its precision measurement goals and has directly informed near detector design choices as described in Section~\ref{sect:design}.

\subsection{Low-Energy Physics}
The DUNE far detector is sensitive to neutrinos 
%with energies in the range up to about 100~MeV, such as those 
produced by the Sun and in core-collapse supernovae with energies in the range $\sim$5-100~MeV. Charged-current interactions of neutrinos from around 5~MeV to several tens of MeV create short electron tracks in liquid argon, potentially accompanied by gamma-ray and other secondary particle signatures.
 
 Core-collapse supernovae within our own Galaxy are expected to occur once every few decades; there is a reasonable chance for one to occur during the several-decade expected lifetime of the DUNE experiment. Because these events are so rare, it is critical that experiments be prepared to capture as much data as possible when one does occur. This need largely drives DUNE's requirements for detector livetime, DAQ and triggering systems, and reconstruction capabilities for low-energy events. 
 DUNE's expected energy threshold is a few MeV of deposited energy and the expected energy resolution is around 10-20\% for energies in the few tens of MeV range. While the expected event rate varies significantly among models of supernova bursts, the 40-kt (fiducial) DUNE detector would be expected to observe approximately 3000 neutrinos from a supernova burst at 10~kpc. 
 Argon detectors have unique sensitivity to electron neutrinos as opposed to antineutrinos;
 because DUNE's far detector is a liquid-argon TPC, the dominant interaction is charged-current absorption of $\nue$ on \argon40: ~$\nue +$\argon40 $\rightarrow e^{-} +$\K40$^{*}$. 
 %This sensitivity to electron neutrinos, as opposed to antineutrinos, is unique to argon detectors.  
 Subdominant $\bar{\nu}_e$ charged-current and neutrino-electron elastic scattering will also contribute.  Another mode of interest is neutral-current scattering on argon nuclei, which may be identified by a cascade of de-excitation gammas. DUNE will participate in worldwide multi-messenger astronomy efforts, such as SNEWS \cite{Antonioli:2004zb}, and, will be able to provide pointing information to localize an observed supernova~\cite{Abi:2020evt,ajpointingtalk}.
 
\begin{figure}[htbp]
 \centering
\includegraphics[width=3in]{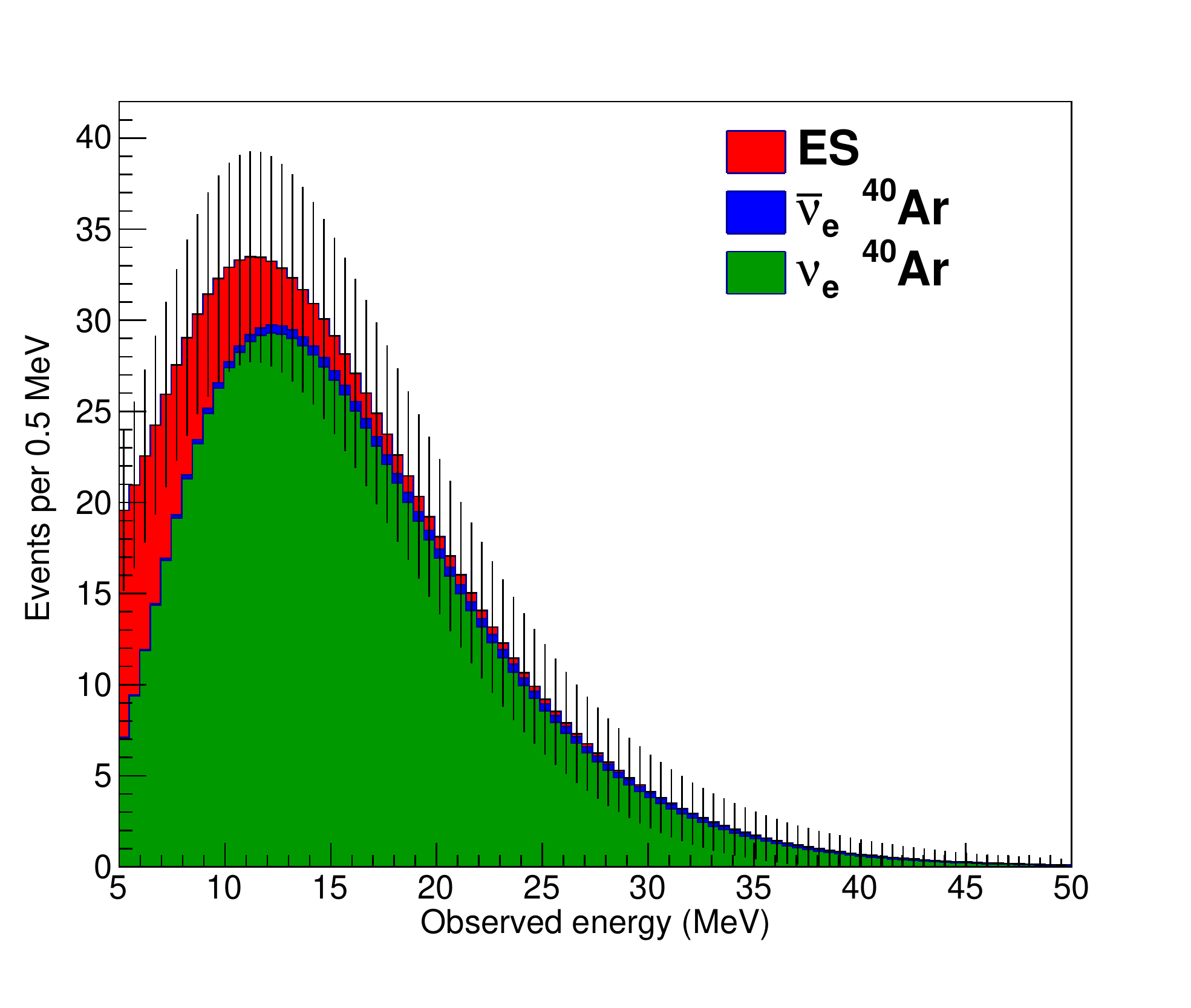}
	\includegraphics[width=2.8in]{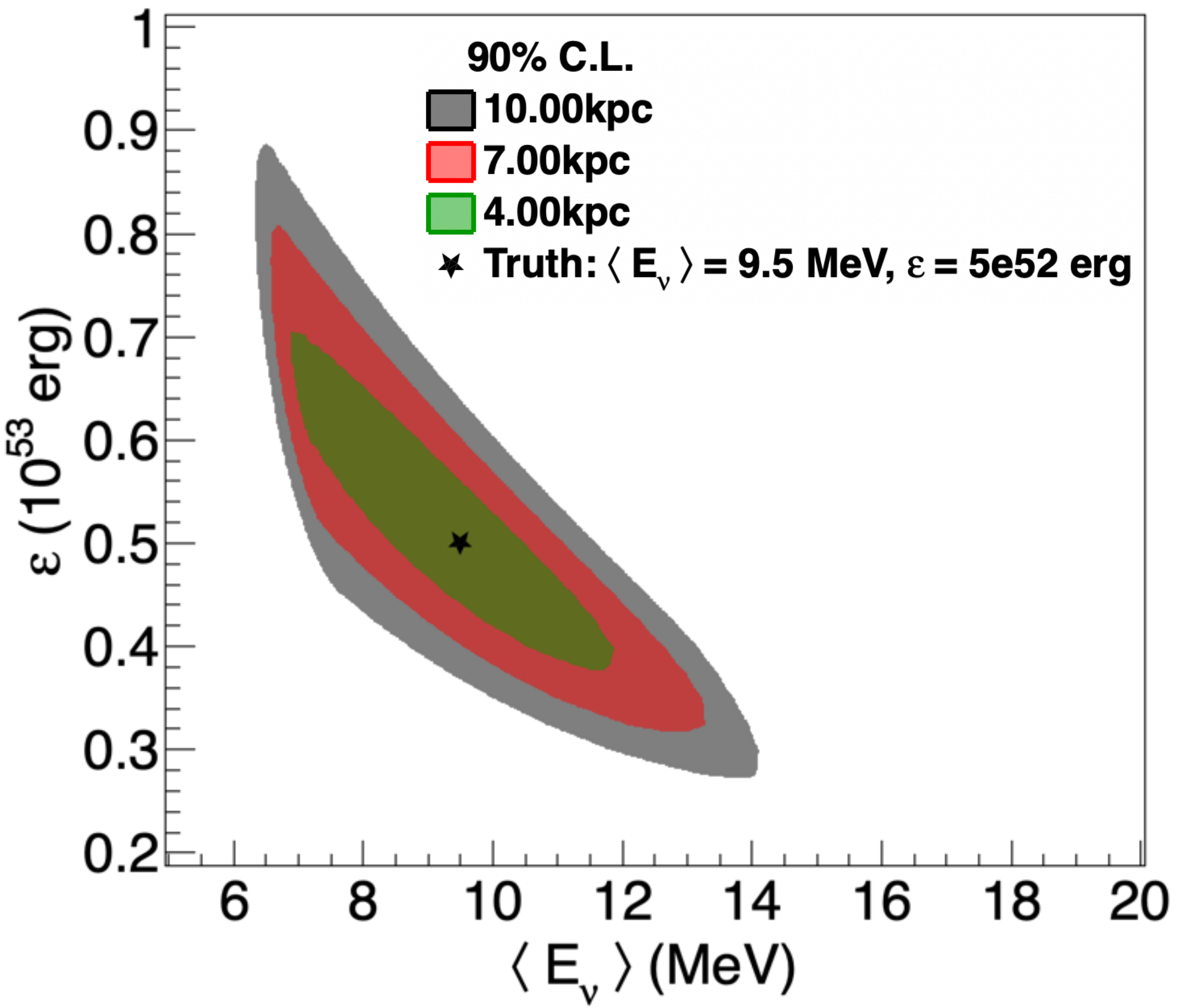}
\caption{Left: Expected measured spectrum in DUNE, for the full 40-kt fiducial mass far detector, as a function of observed energy, after detector response smearing and integrated over time, for the model in~\cite{Hudepohl:2013zsj} with no flavor transformation.  Event rates are computed using \snowglobes~\cite{snowglobes} with a transfer matrix based on full DUNE simulation and reconstruction, and shown for neutrino-electron elastic scattering (ES), as well as neutrino- and antineutrino-nucleus scattering.  Right: Sensitivity regions in $(\langle E_\nu \rangle, \varepsilon)$ space (profiled over pinching parameter $\alpha$) for the $\nu_e$ spectrum
    for three different supernova distances, for the full 40-kt fiducial mass far detector.  
    \snowglobes~assumes a
   transfer matrix made using MARLEY~\cite{PhysRevC.103.044604} with a 20\% Gaussian resolution on detected energy, and a step efficiency function with a 5 MeV
    detected energy threshold. \label{fig:garchingspec}}
\end{figure}

 In a core-collapse supernova, the neutrino signal starts with a short, sharp ``neutronization" burst primarily composed of $\nue$. This is followed by an ``accretion" phase lasting several hundred milliseconds, and then a ``cooling" phase which lasts about 10 seconds and represents the bulk of the signal, roughly equally divided among all flavors of neutrinos and antineutrinos. The flavor content and spectra of neutrinos change throughout these phases, so the supernova's evolution can be mapped out using the neutrino signal.  Information about the progenitor, the collapse, the explosion, and the remnant, as well as information about neutrino properties, are contained in this signal. The flux spectrum may be parameterized by the ``pinched-thermal" model\cite{Tamborra:2012ac}.  DUNE will have sensitivity to determining the parameters describing the $\nu_e$ spectrum; see Fig.~\ref{fig:garchingspec}.
 Other astrophysical observables include: the formation of a black hole, which would cause a sharp signal cutoff
  (e.g.,~\cite{Beacom:2000qy,Fischer:2008rh,Li:2020ujl}); shock wave effects (e.g.,~\cite{Schirato:2002tg}), which would cause a
  time-dependent change in flavor and spectral composition as the
  shock wave propagates; the standing accretion shock instability
  (SASI)~\cite{Hanke:2011jf,Hanke:2013ena}, a ``sloshing'' mode
  predicted by three-dimensional neutrino-hydrodynamics simulations of supernova cores which would give an oscillatory flavor-dependent
  modulation of the flux; and turbulence effects~\cite{Friedland:2006ta,Lund:2013uta}, which would
  also cause flavor-dependent spectral modification as a function of
  time.

 Study of the energy balance of a supernova burst can provide constraints on new physics scenarios, the existence of which would alter the energy transport process within the explosion; the existing data from SN1987 can already begin to constrain some scenarios. Neutrino flavor transformation probabilities are complex in a supernova burst because of the kinematics of the explosion itself and the possibility for neutrino-neutrino  scattering and collective modes of oscillation. These effects will imprint on the neutrino signal and can be used to study these phenomena experimentally. The true neutrino mass ordering has a strong impact on the expected signal~\cite{Scholberg:2017czd}, particularly in early times including the neutronization burst. Knowledge of the mass ordering from DUNE's own long-baseline oscillation measurements may be used to better extract other particle and astrophysical knowledge from the observed supernova burst signal.
 
 Neutrinos and antineutrinos from other astrophysical sources, such as solar~\cite{Capozzi:2018dat} and diffuse background supernova neutrinos~\cite{Moller:2018kpn}, are also potentially detectable. While detection of these sources will be challenging, particularly because of the presence of radioactive background in the detector, initial studies suggest potential for DUNE to select a sample of solar neutrinos that would allow a significant improvement in the  measurement of $\dm{21}$ as well as observations of the $hep$ and $^8$B solar neutrino flux. Development of reconstruction, calibration, and triggering/DAQ infrastructure will play an important role in enabling a broader physics program at low energies.
 
 \subsection{Physics Beyond the Standard Model}
 The deep underground location of the DUNE far detector facilitates sensitivity to nucleon decay and other rare processes. DUNE can probe a rich and diverse BSM phenomenology including searches for dark matter, sterile neutrino mixing, nonstandard neutrino interactions, CPT violation, new physics enhancing neutrino trident production, and baryon number violating processes. 
 
 Experimental results in tension with the three-neutrino-flavor paradigm, which may be interpreted as mixing between the known active neutrinos and one or more sterile states, have led to a rich and diverse program of searches for oscillations into sterile neutrinos~\cite{ref:tension,Gariazzo:2017fdh}. DUNE is sensitive over a broad range of potential sterile neutrino mass splittings by looking for disappearance of charged-current and neutral-current neutrino interactions over the long distance separating the ND and FD, as well as over the short baseline of the ND. With a longer baseline, a more intense beam, and a high-resolution large-mass FD, compared to previous experiments, DUNE provides a unique opportunity to improve significantly on the sensitivities of the existing probes, and greatly enhance the ability to map the extended parameter space if a sterile neutrino is discovered. A generic characteristic of most models explaining the neutrino mass
pattern is the presence of heavy neutrino states beyond the
three light states of the Standard Model of particle
physics~\cite{Mohapatra:1998rq,Valle:2015pba,Fukugita:2003en}. These
types of models, as well as those of light sterile neutrinos, imply that the $3 \times 3$ PMNS matrix is not unitary due to mixing with additional states. DUNE can constrain the parameters describing non-unitarity~\cite{Blennow:2016jkn,Escrihuela:2016ube} with precision comparable to that from present oscillation experiments.

Non-standard interactions (NSI), affecting neutrino propagation through the Earth, can significantly modify the data to be collected by DUNE as long as the new physics parameters are large enough~\cite{Farzan:2017xzy}. Leveraging its very long baseline and wide-band beam, DUNE is uniquely sensitive to these probes. DUNE can substantially improve the bounds on, for example, the NSI parameter $\varepsilon_{\tau\tau}-\varepsilon_{\mu\mu}$ and the non-diagonal NSI parameters.

CPT symmetry is a cornerstone of model-building in modern physics. 
Using beam neutrinos, DUNE can improve the present limits on Lorentz and CPT violation by several orders of magnitude~\cite{Streater:1989vi,Barenboim:2002tz,Kostelecky:2003cr,Diaz:2009qk,Kostelecky:2011gq,Barenboim:2017ewj,Barenboim:2018lpo,Barenboim:2018ctx}, a very important test of these fundamental assumptions underlying quantum field theory.
Atmospheric neutrinos are a unique tool for studying neutrino oscillations: the oscillated flux contains all flavors of neutrinos and antineutrinos, is very sensitive to matter effects and to both \dm{} parameters, and covers a wide range of $L/E$. Studying atmospheric neutrinos in DUNE is a promising approach
to search for BSM effects such as Lorentz and CPT violation.

Neutrino trident production is a weak process in which a neutrino, scattering off the Coulomb field of a heavy nucleus, generates a pair of charged leptons~\cite{Czyz:1964zz,Lovseth:1971vv,Fujikawa:1971nx,Koike:1971tu,Koike:1971vg,Brown:1973ih,Belusevic:1987cw,Zhou:2019vxt,Zhou:2019frk}. The high-intensity muon-neutrino flux at the DUNE near detector will lead to a sizable production rate of trident events, offering excellent prospects to improve ~\cite{Altmannshofer:2019zhy,Ballett:2018uuc,Ballett:2019xoj} on existing measurements. A deviation from the event rate predicted by the Standard Model could be an indication of new interactions mediated by new gauge bosons~\cite{Altmannshofer:2014pba}. 

DUNE will carry out an extensive program of dark sector particle searches at the ND, including searches for axion-like particles and low-mass dark matter (LDM)~\cite{Romeri_2020}. This is possible because the ND is close enough to the intense beam source to sample a substantial dark matter (DM) flux, assuming that DM is produced. The boosted nature of LDM produced in the beamline enables DUNE to access a mass range that is not reachable by existing direct search experiments due to intrinsic radioactive impurities. In addition, the PRISM capability of the ND further enhances the search and provides additional control of Standard Model neutrino backgrounds. A discovery of DM could guide the design of a DM beam and a detailed study of the DM properties. In Phase II, ND-GAr adds additional unique sensitivity to heavy neutral leptons (HNL) and heavy axions, where the signature is the decay of a new particle in the detector. In such searches, the signal rate scales with the detector volume while backgrounds from neutrino interactions scale with detector mass, so the low density of ND-GAr is ideal.

It is also possible that boosted dark matter (BDM) particles are created in the universe under non-minimal dark-sector scenarios~\cite{Agashe:2014yua,Belanger:2011ww}, and can reach terrestrial detectors. The DUNE far detector is expected to possess competitive sensitivity to BDM signals from various sources in the current universe such as the galactic halo~\cite{Agashe:2014yua,Alhazmi:2016qcs,Kim:2016zjx,Giudice:2017zke,Chatterjee:2018mej,Kim:2018veo,Necib:2016aez}, the sun~\cite{Alhazmi:2016qcs,Kim:2018veo,Kong:2014mia,Huang:2013xfa,Berger:2014sqa, Berger:2019ttc}, and dwarf spheroidal galaxies~\cite{Necib:2016aez}. One particular signature that is very distinct in the DUNE FD is low-mass BDM interacting inelastically, and upscattering to a higher-mass state, which subsequently decays to low mass DM. This is expected to produce two leptons in the initial interaction, and either an additional lepton or a nucleon recoil some distance away. DUNE's sensitivity in this channel covers a mass range that has not previously been probed.

The excellent imaging, as well as calorimetric and particle identification capabilities, of the LArTPC technology implemented for the DUNE far detector will facilitate searches for a broad range of baryon-number violating processes. Reconstruction of these events, which have final-state particle kinetic energy of order 100 MeV, is a significant challenge, made more difficult by final state interactions (FSI), which generally reduce the energy of observable particles. The dominant background for these searches is from atmospheric neutrino interactions. For example, a muon from an atmospheric $\numu n \rightarrow \mu^{-}p$ interaction may be indistinguishable from a muon from $K^{+}\rightarrow\mu^{+}\rightarrow e^{+}$ decay chain from \ptoknubar decay, such that identification of the event relies on the kaon-proton discrimination. 

%Figure\ref{fig:pida} shows the particle identification performance for kaons from  proton  decay, muons from  kaon decay, and protons produced by atmospheric neutrino interactions.

%\begin{figure}
%\centering
%\includegraphics[width=0.5\columnwidth]{PIDA.pdf}
%\caption{Particle identification using $PIDA$ for muons and kaons in simulated proton decay events, \ptoknubar, and protons in simulated atmospheric neutrino background events.  The curves are normalized by area.}
%\label{fig:pida}
%\end{figure}

Sensitivity to several of these processes has been studied using the full DUNE simulation and reconstruction analysis chain, including the impact of nuclear modeling and FSI on a BDT-based selection algorithm. 
With an expected 30\% signal efficiency, including anticipated reconstruction advances, and an expected background of one event per Mt-yr, a 90\% C.L. lower limit on the proton lifetime in the \ptoknubar channel of $1.3 \times 10^{34}$~years can be set, assuming no signal is observed for a 400 kt-yr exposure.
Another potential mode for a baryon number violation search is the decay of the neutron into a charged lepton plus meson, i.e.,~\ntoek. The lifetime sensitivity for a 400 kt-yr exposure is estimated to be $1.1 \times 10^{34}$~years.
Neutron-antineutron (\nnbar) oscillation is a $|\Delta B|=2$ process, which also leads to $\Delta B=-2$ dinucleon decays. This type of baryon number violation has never been observed but is predicted to occur by many BSM theories\cite{addazinnbar,PHILLIPS20161}, and has been searched for in free neutron propagation experiments and nucleon decay detectors. The expected limit for the oscillation time of free neutrons for a 400 kt-yr DUNE exposure is calculated to be $5.5 \times 10^{8}$~s.

%And here is an example of a table: %Table~\ref{tab:oscpar_nufit}.

%\begin{table}[htbp]
%    \centering
%    \begin{tabular}{lcc}
%      \hline
% Parameter &    Central value & Relative uncertainty %\\
%\hline\hline
%$\theta_{12}$ & 0.5903 & 2.3\% \\ \hline
%$\theta_{23}$ (NO) & 0.866  & 4.1\% \\ 
%$\theta_{23}$ (IO) & 0.869  & 4.0\% \\ \hline
%$\theta_{13}$ (NO) & 0.150  & 1.5\% \\ 
%$\theta_{13}$ (IO) & 0.151  & 1.5\% \\ \hline
%$\Delta m^2_{21}$ & 7.39$\times10^{-5}$~eV$^2$ & %2.8\% \\ \hline
%$\Delta m^2_{32}$ (NO) & 2.451$\times10^{-3}$~eV$^2$ %&  1.3\% \\
%$\Delta m^2_{32}$ (IO) & -2.512$\times10^{-3}$~eV$^2$ %&  1.3\% \\
%\hline
%$\rho$ & 2.848 g cm$^{-3}$ & 2\% \\
%\hline
%    \end{tabular}
%    \caption[Parameter values and uncertainties from a global fit to neutrino oscillation data]{Central value and relative uncertainty of neutrino oscillation parameters from a global fit~\cite{Esteban:2018azc,nufitweb} to neutrino oscillation data. The matter density is taken from Ref.~\cite{Roe:2017zdw}. Because the probability distributions are somewhat non-Gaussian (particularly for $\theta_{23}$), the relative uncertainty is computed using 1/6 of the 3$\sigma$ allowed range from the fit, rather than 1/2 of the 1$\sigma$ range. For $\theta_{23}$, $\theta_{13}$, and $\Delta m^2_{31}$, the best-fit values and uncertainties depend on whether normal mass ordering (NO) or inverted mass ordering (IO) is assumed.}
%    \label{tab:oscpar_nufit}
%\end{table}

%% file: sections/practical.tex
\section{Practical Considerations}
\label{sect:practical}

The construction of LBNF and DUNE is resource-limited. The highest priority is to deliver the full LBNF facility, including far site conventional facilities to accommodate four 17-kt FD (each at least 10-kt fiducial) modules, near site conventional facilities to support the full ND, and a 1.2 MW beam that can be upgraded to 2.4 MW. DUNE will be built in two phases, detailed in Table~\ref{tab:phases}. In the most recent funding profile scenarios, DUNE Phase I is anticipated to begin collecting physics data with the far detector in the late 2020s, with the beam and near detector coming online by 2032. The Phase I configuration is sufficient for early physics goals, and is capable of producing world-class results including the determination of the neutrino mass ordering. However, Phase I can not achieve the precision neutrino oscillation physics goals laid out in the P5~\cite{HEPAPSubcommittee:2014bsm}; the full scope is critical for both the oscillation and non-oscillation physics programs of DUNE.
In this section, we describe the physics reach of each phase of the experiment and the impact of the staging scenarios that arise from funding constraints on the DUNE timeline. We quantify the importance of realizing the full DUNE experiment design (Phase II), including a 40-kt fiducial far detector, a 2.4 MW beam, and the full near detector. 
%We emphasize that DUNE's physics program is in significant jeopardy if the full scope of the design is not realized and call on the community and P5 to reaffirm that realization of the full DUNE design is the highest priority for the coming decade.

\begin{table}[h]
    \centering
    \begin{tabular}{c|c|c|c}
        Parameter  & Phase I      & Phase II & Impact \\ \hline
        FD mass    & 20 kt fiducial       & 40 kt fiducial    & FD statistics \\
        Beam power & up to 1.2 MW & 2.4 MW   & FD statistics \\
        ND config  & ND-LAr,TMS, SAND          & ND-LAr, ND-GAr, SAND   & Syst. constraints \\
    \end{tabular}
    \caption{A description of the two-phased approach to DUNE. ND-LAr, including the PRISM movement capability, and SAND are present in both phases of the ND.}
    \label{tab:phases}
\end{table}

% Is this paragraph necessary?
%Many DUNE sensitivity studies have been presented with exposure quoted in kt-MW-years, such that the details of funding availability and staging scenarios do not have to be considered. The sensitivities presented in the FD TDR~\cite{tdr-vol-2} assume a technically-limited schedule where the full FD fiducial mass is attained by the third year of operation, and the beam is upgraded to 2.4 MW after six years. The practicalities of balancing funding among HEP activities are such that achieving this technically limited schedule is not feasible. 

%There are three aspects of the experiment where delays due to funding limitations affect DUNE's physics sensitivities:

%\begin{enumerate}
%    \item Far Detector mass (e.g. delays to module 3 and 4 installation),
%    \item Beam power (e.g. delay of the upgrade to 2.4 MW),
%    \item Near detector capability (e.g. delay in installing ND-GAr).
%\end{enumerate}

All three of the experiment parameters listed in Table~\ref{tab:phases} (far dectector mass, beam power, and near detector configuration) affect DUNE's oscillation physics sensitivities. In the case of the FD mass and beam power, Phase II is required in order to attain sufficient far detector statistics to reach the required precision. Going from Phase I to Phase II quadruples the FD event rate, such that exposure (in kt-MW-years) accumulates four times faster. The ND constrains systematic uncertainty in predictions of the expected FD event spectrum, which would otherwise limit precision measurements of oscillation parameters. In addition to the oscillation physics, the far detector mass also affects the low-energy physics program (e.g. supernova and solar neutrinos) as well as searches for rare signals from natural sources (e.g. nucleon decays), the beam power affects searches for BSM signals that come from the beam (e.g. heavy neutral leptons, neutrino tridents, etc.), and the ND plays a key role in some BSM searches and will faciliate a broad suite of neutrino cross-section and Standard Model parameter measurements.

% More about why we need ND-LAr here
%ETW move to ND section

\begin{figure}[htbp]
  \centering
  \includegraphics[width=0.45\linewidth]{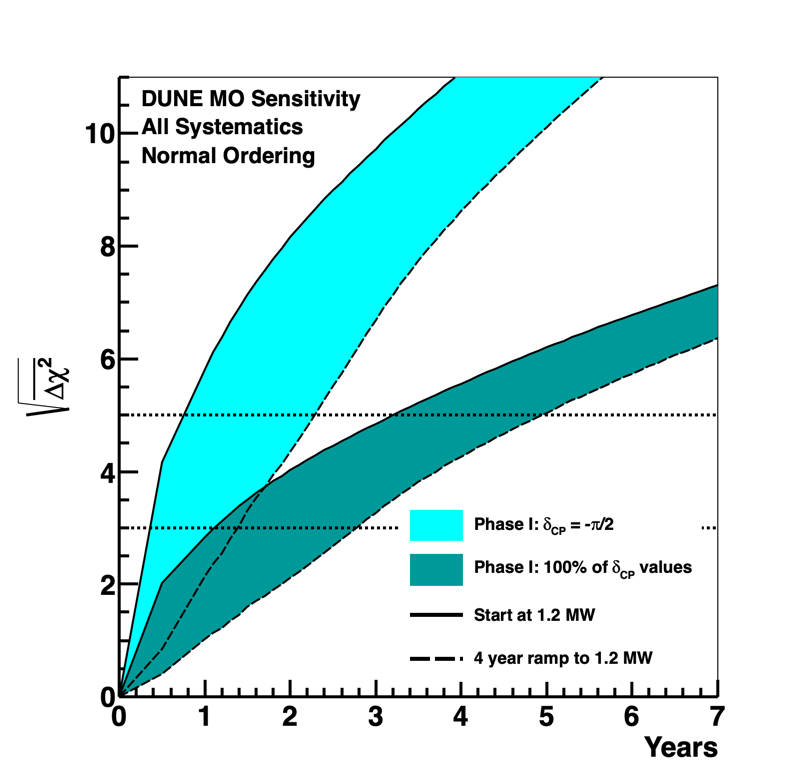}
  \includegraphics[width=0.45\linewidth]{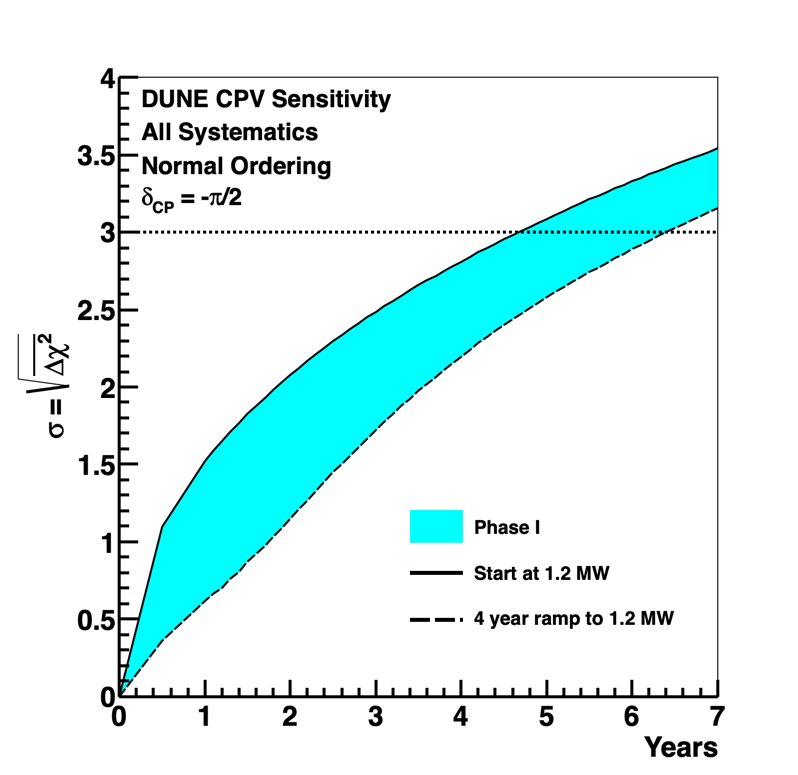}
  \caption{Sensitivity to the neutrino mass ordering (left) and CP violation for $\deltacp = -\pi/2$ (right) in Phase~I. The cyan bands show the sensitivity if $\deltacp = -\pi/2$ and the green band in the left plot shows the sensitivity for 100\% of $\deltacp$ values. The width of the bands shows the impact of potential beam power ramp up; the solid upper curve is the sensitivity if data collection begins with 1.2 MW beam power and the lower dashed curve shows a conservative beam ramp scenario where the full power is achieved after 4 years.}
    \label{fig:sens_phase1}
\end{figure}

Figure~\ref{fig:sens_phase1} shows the sensitivity to the neutrino mass ordering and CP violation that can be achieved with the Phase I configuration. The mass ordering can be unambiguously determined within the first two years of the experiment if $\deltacp = -\pi/2$ and within 3-5 years for any value of $\deltacp$. The Phase I detector configuration will also have 3$\sigma$ sensitivity to CP violation if $\deltacp = -\pi/2$. The width of the bands corresponds to the beam turn-on; the bottom curve assumes the currently-understood ramp-up schedule and the top curve assumes that the beam turns on at 1.2 MW on day one; the most likely true scenario falls within the band. The exposure and time required to reach these milestones are summarized in Table~\ref{tab:sens_milestones}.
In light of expected results from contemporary experiments, such as JUNO\cite{An:2015jdp} and Hyper-Kamiokande\cite{Abe:2015zbg}, 
%and the need to maintain engagement, enthusiasm, and partner funding, 
it is important to achieve the full 1.2 MW proton beam power, and thus interesting physics results, as early as possible. However, even in the least favorable case, unambiguous determination of the neutrino mass ordering is possible in Phase I.

\begin{table}[h]
    \centering
    \begin{tabular}{llcc}
         Experiment Stage \hspace{10pt}  & Physics Milestone &Exposure  & Years  \\
         &  & (kt-MW-years) & (Staged) \\ \hline
         Phase I & 5$\sigma$ MO ($\deltacp = -\pi/2$) & 16 & 1-2\\
         & 5$\sigma$ MO (100\% of $\deltacp$ values) & 66 & 3-5\\
         & 3$\sigma$ CPV ($\deltacp = -\pi/2$) & 100 & 4-6\\ \hline
         Phase II & 5$\sigma$ CPV ($\deltacp = -\pi/2$) & 334 & 7-8\\
         & $\deltacp$ resolution of 10 degrees ($\deltacp = 0$) & 400 & 8-9\\
         & 5$\sigma$ CPV (50\% of $\deltacp$ values) & 646 & 11 \\
         & 3$\sigma$ CPV (75\% of $\deltacp$ values) & 936 & 14\\
         & sin$^22\theta_{13}$ resolution of 0.004 & 1079 & 16 \\
    \end{tabular}
    \caption{Exposure, in kt-MW-years, and time, in calendar years, required to reach selected physics milestones. The time in years assumes that Phase I is complete at Year 0 and that the Phase II staging scenario described in the text is realized. The range of time in years covers the effect of the beam ramp, with the lower bound corresponding to full 1.2 MW proton beam power at Year 0 and the higher bound corresponding to a scenario where the full power is achieved after 4 years. When no range is provided, the difference between these scenarios is less than one year. Time in years is rounded to the nearest whole year.}
    \label{tab:sens_milestones}
\end{table}

\begin{figure}[htbp]
  \centering
  \includegraphics[width=0.9\linewidth]{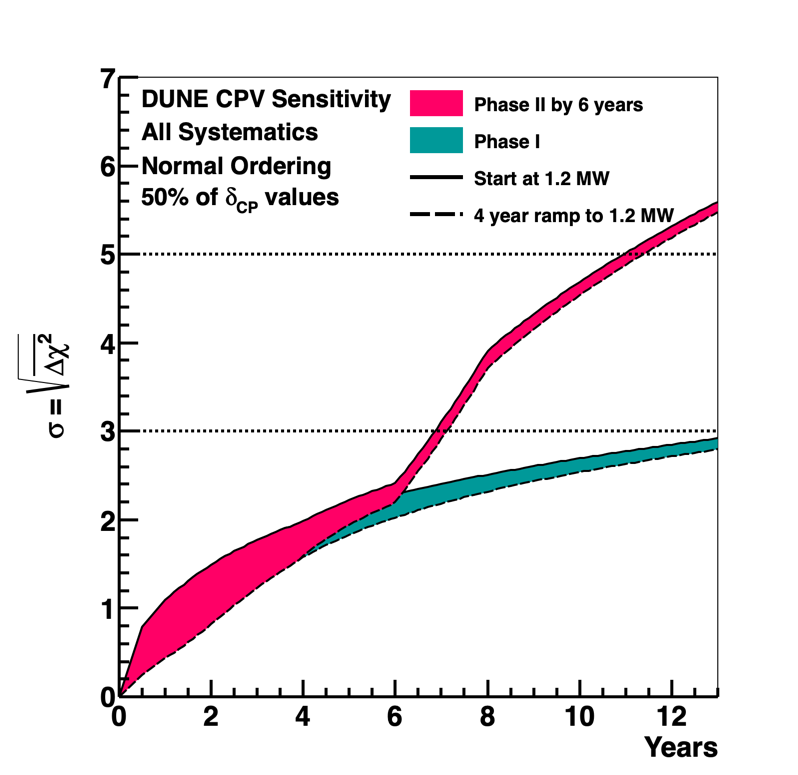}
  \caption{Sensitivity to CP violation for 
  %$\deltacp = -\pi/2$ (left) and for 
  50\% of $\deltacp$ values %(right) 
  in Phase I (green band) and 
  in a scenario where Phase II is achieved after 6 years (red band). The width of the bands shows the impact of potential beam power ramp up; the solid upper curve is the sensitivity if data collection begins with 1.2 MW beam power and the lower dashed curve shows a conservative beam ramp scenario where the full power is achieved after 4 years.}
    \label{fig:newplot}
\end{figure}

Figure~\ref{fig:newplot} shows the sensitivity 
%to maximal CP violation and 
to CP violation over 50\% of $\deltacp$ values that can be achieved with the Phase II configuration. The green band represents the Phase I configuration as seen in Fig.~\ref{fig:sens_phase1}: 1.2 MW beam, 2 FD modules, and a limited ND. The red band represents a scenario where upgrades are pursued aggressively: an upgrade to 2.4 MW after 6 years of operation, the third and fourth FD module added in years 4 and 6, respectively, and the ND upgraded to include ND-GAr after 6 years. The width of the band corresponds to the beam turn-on as in Fig.~\ref{fig:sens_phase1}. The exposure and time required to achieve selected physics milestones are summarized in Table~\ref{tab:sens_milestones}.
%Sensitivity to maximal CPV at $3\sigma$ is achieved in the Phase I program, but 
Phase II is required to reach high significance over a broad range of $\deltacp$ values. CP violation sensitivity for a broad range of values is directly related to and can be used as a proxy for the precision measurements of $\deltacp$ and other oscillation parameters, which require Phase II. 
%The neutrino mass ordering is determined in Phase I regardless of the true values of other parameters.

\begin{figure}[htbp]
  \centering
  \includegraphics[width=0.32\linewidth]{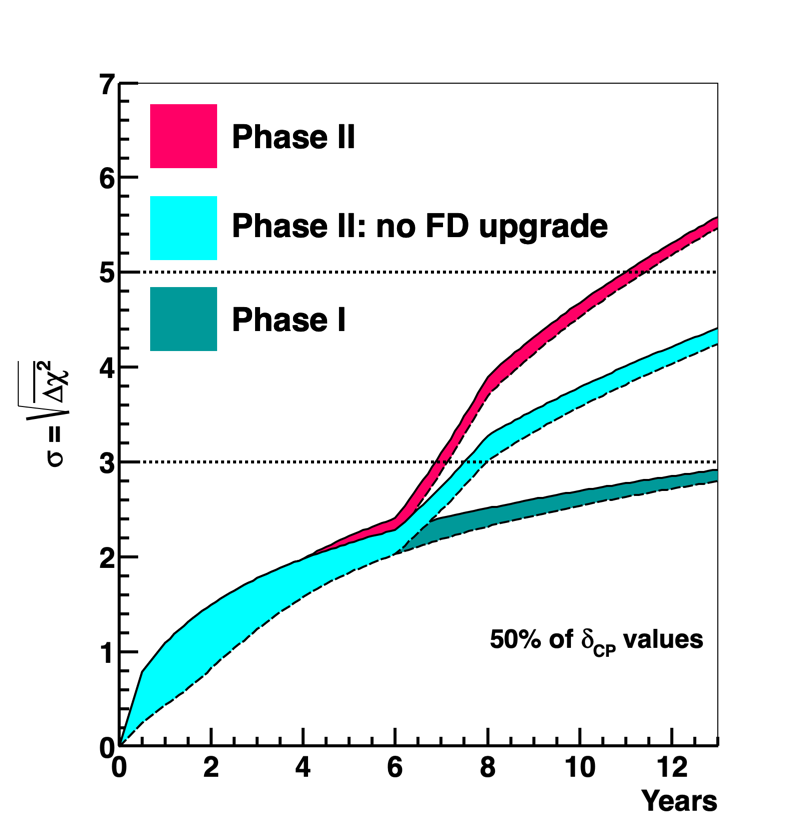}
  \includegraphics[width=0.32\linewidth]{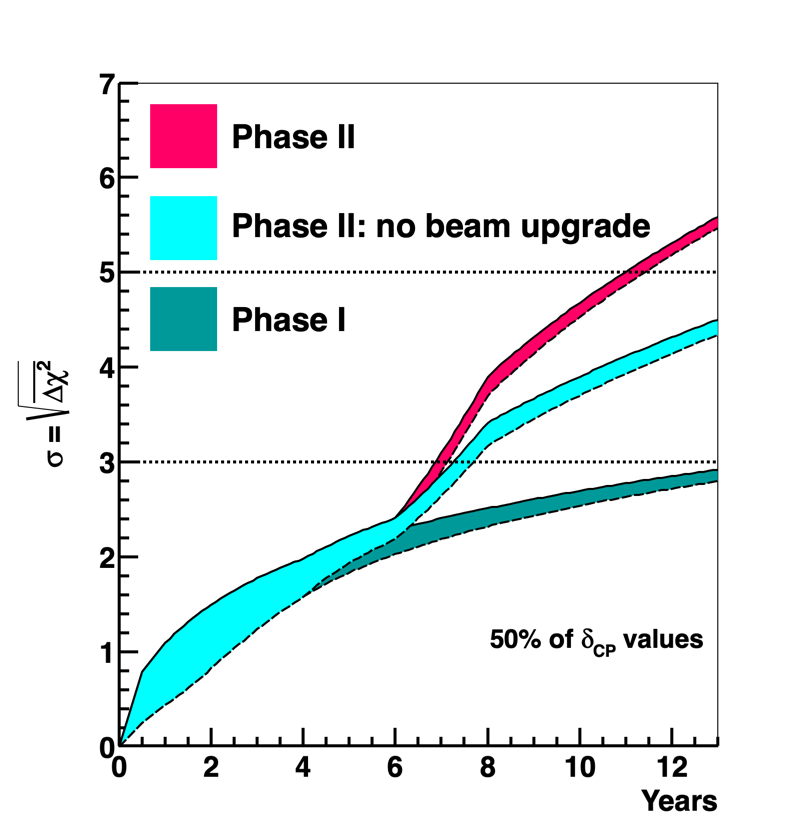}
  \includegraphics[width=0.32\linewidth]{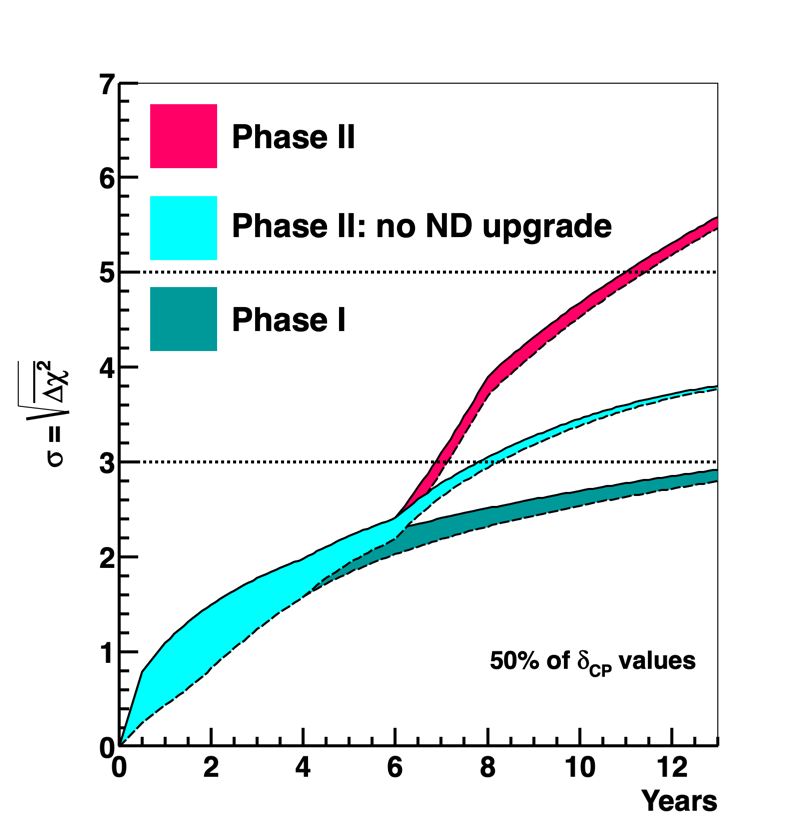}
  \caption{Sensitivity to CP violation for 50\% of $\deltacp$ values, as a function of time in calendar years. The width of the bands shows the impact of potential beam power ramp up; the solid upper curve is the sensitivity if data collection begins with 1.2 MW beam power and the lower dashed curve shows a conservative beam ramp scenario where the full power is achieved after 4 years. The green bands show the Phase I sensitivity and the red bands shows the Phase II sensitivity. In each plot the cyan band shows the Phase II sensitivity if one of the three upgrades does not occur.
  The left plot shows the sensitivity without the FD upgrade, the middle plot shows the sensitivity without the beam upgrade, and the right plot shows the sensitivity without the ND upgrade, illustrating that each is necessary to achieve DUNE's physics goals.}
    \label{fig:nminusone}
\end{figure}

The impact of the far detector mass, beam power, and near detector upgrades is illustrated in Figure~\ref{fig:nminusone}, which shows the sensitivity to CP violation for 50\% of $\deltacp$ values in the absence of each of the three upgrades, compared to the Phase I and Phase II sensitivity. As seen in Figure~\ref{fig:nminusone}, all three upgrades are necessary to achieve DUNE's precision physics goals. The impact of FD mass, beam power, and the ND are described further in Sections~\ref{sec:fd_mass}, \ref{sec:beam_power}, and \ref{sec:nd}, respectively. 
%The impact of the beam ramp-up to 1.2 MW is described in Section~\ref{sec:beam_ramp}.

\subsection{Impact of Far Detector Fiducial Mass}
\label{sec:fd_mass}

% reasonable amount of time --> 10-year timescale? ETW - we don't quite make it within 10 years even with the upgrades so maybe say 12 years?
The far detector event rate is directly proportional to the far detector fiducial mass, such that there is a direct correlation between fiducial mass and sensitivity until the experiment becomes systematics limited. 
%at very large exposure. 
The first and second FD modules (FD-1 and FD-2) are part of the Phase I project, while the third and fourth modules (FD-3 and FD-4) are part of Phase II. The left plot in Figure~\ref{fig:nminusone} shows the impact of FD fiducial mass on sensitivity to CP violation over 50\% of $\deltacp$ values, as an example of the precision oscillation physics program. 
%The individual impact of adding FD-3 and FD-4 to Phase I is shown on the left and the Phase II sensitivity without FD-3 and FD-4 is shown on the right. 
The installation of FD-3 and FD-4 are necessary, but not sufficient, to achieve DUNE's precision oscillation physics goals; without these detectors, DUNE can not achieve 5$\sigma$ sensitivity to CP violation for 50\% of $\deltacp$ values within a 12 year timescale.

%\begin{figure}[htbp]
%  \centering
%  \includegraphics[width=0.45\linewidth]{graphics/cpv_exp_staging_2022wp_comparemassupgrades_cpv50_alt.png}
%  \includegraphics[width=0.45\linewidth]{graphics/cpv_exp_staging_2022wp_beamupgrade_ndgar_cpv50_alt.png}
%  \caption{Sensitivity to CP violation for 50\% of $\deltacp$ values, as a function of time in calendar years. The width of the bands shows the impact of a conservative beam ramp scenario where the full power is achieved after 4 years. The green curve shows the Phase I sensitivity. In the left plot, the red curve shows the sensitivity when FD-3 and FD-4 are added after four and six years, respectively. In the right plot, the red curve shows the Phase II sensitivity without FD-3 and FD-4, illustrating that this upgrade is necessary to reach DUNE's goal sensitivity.}
%    \label{fig:fdmass}
%\end{figure}

Among the Phase II upgrades, FD-3 and FD-4 have the greatest impact on the broader DUNE physics program because they affect searches for signals from natural sources. %Figure~\ref{fig:fdmass_snb} shows the time and energy profile of the low-energy neutrinos from a galactic supernova burst. 
The additional FD mass doubles the number of observed neutrinos. For a galactic SNB, the number of neutrinos is large in either case, but the additional statistics are extremely useful for measurements that are sensitive to the astrophysics of the core collapse. The additional mass is even more relevant for supernova bursts in the Large or Small Magellanic Cloud. DUNE's sensitivity to other rare signals such as nucleon decays is dependent on the exposure, measured in kt-years, and scales linearly with the FD mass.

\subsection{Impact of Beam Power}
\label{sec:beam_power}

For the accelerator-based physics program including neutrino oscillations, the impact of beam power is exactly parallel to that of far detector fiducial mass: the far detector event rate is directly proportional to the proton beam power. This can be seen in Figure~\ref{fig:nminusone}, 
where the left and middle plots, illustrating the Phase II sensitivity without the FD and beam upgrades, respectively, are nearly identical. 
The beam power upgrade is necessary, but not sufficient, to achieve DUNE's precision oscillation physics goals; without 2.4 MW proton beam power, DUNE can not achieve 5$\sigma$ sensitivity to CP violation for 50\% of $\deltacp$ values within a 12 year timescale. In PIP-II~\cite{pip2-2013} will provide the 1.2~MW proton beam required for Phase I. The upgrade to 2.4~MW is beyond the scope of PIP-II; it requires replacement of the Booster and possible upgrades of the Main Injector. An upgrade path has been proposed~\cite{Ainsworth:2021ahm} that could meet the needs of DUNE and address much of the science program presented by the Booster Replacement Science Working Group~\cite{Arrington:2022pon}.

\subsection{Impact of Near Detector}
\label{sec:nd}

DUNE is expected to begin taking data with a limited near detector. The DUNE Phase I ND design consists of ND-LAr, TMS, and SAND. The core requirement of the ND is to inform predictions of the reconstructed neutrino energy spectrum in the FD, which is determined by adding the lepton and visible hadronic energy. The hadronic energy in particular is sensitive to neutrino cross sections and LAr TPC detector response in a complex way. As such, the most critical ND component for both Phase I and Phase II is ND-LAr, which is the only detector that can measure neutrino-argon interactions in a way that can be directly extrapolated to the FD.

%To reduce costs, ND-LAr is not large enough to measure most forward muons by range, and must be functionally coupled to a downstream spectrometer. ND-LAr and a capable muon catcher are required for DUNE to produce any credible oscillation physics measurement. Therefore the Phase I near detector consists of ND-LAr and the Temporary Muon Spectrometer (TMS), which move up to 30.5~m off-axis. TMS serves only to measure muons produced in ND-LAr that exit the back of the detector; it is not expected that neutrino interactions in TMS itself will improve constraints. An on-axis detector provides a safeguard against changes in the beam that occur while ND-LAr and TMS are off-axis. SAND fulfills this role, and provides additional neutrino cross section constraints on several nuclear targets that enhance the ND physics program. 

In Phase I, ND-LAr is supported by TMS to measure the charge and momentum of forward muons. This initial ND configuration is sufficient for the needs of the experiment for early exposures up to about 200 kt-MW-yrs, where the primary oscillation physics milestones are the determination of the mass ordering and sensitivity to maximal CPV. 
For exposures less than 100 kt-MW-years, there will be 3-5$\sigma$ sensitivity to the neutrino mass ordering with the Phase I ND. Similarly, in the 100-200 kt-MW-year exposure range, observation of CP violation at the 3$\sigma$ level is possible if $\deltacp=-\pi/2$.

Beyond 200 kt-MW-yrs, systematic uncertainty from inadequacies in the interaction model will limit the ultimate sensitivity of the experiment. 
In Phase II, TMS is replaced with ND-GAr~\cite{garwhitepaper}, a magnetized, high-pressure gaseous argon TPC and calorimeter. The primary impact of ND-GAr is providing high-precision neutrino-argon interaction measurements with $4\pi$ acceptance including sign selection, very low thresholds, and very long hadron interaction lengths that enable exquisite particle ID. These measurements complement ND-LAr and provide the additional precision required to achieve the ultimate oscillation physics goals. ND-GAr also serves as a muon spectrometer for ND-LAr.
Because the beam stability will be directly monitored, constraints on neutrino-argon interactions in ND-GAr can eventually be applied to the full sample, so that FD data taken in Phase I can be combined with FD data from Phase II. After commissioning is complete, ND-GAr will need roughly one year of data in each beam mode to begin to provide constraints.
The right panel of Figure~\ref{fig:nminusone} shows DUNE sensitivity to CP violation for 50\% of $\deltacp$ values in Phase II with and without ND-GAr. Higher significance sensitivity for 50\% or more of $\deltacp$ values eventually becomes impossible without ND-GAr, as do precision measurements of the oscillation parameters that are among the ultimate goals of DUNE. Beyond the neutrino oscillation program, ND-GAr also plays a key role in BSM searches, in particular those in which new particles decay (rather than interact) in the detector volume, as well as in neutrino cross section measurements that are sensitive to nuclear structure.

%% file: sections/snowmass.tex
\section{Message to the Particle Physics Community for Snowmass}
\label{sect:snowmass}

%Copying in the stuff I commented out elsewhere. Need to synthesize into coherent argument.

%Points (not wordsmithed - just notes before doing the real writing)
%-DUNE physics is broad/exciting: Phase I will resolve a long-standing problem in a definitive way that no other experiment can and Phase II has great precision reach. Not just oscillation physics
%-Support must be maintained for Phase I, including ND-LAr (how to justify)
%-Upgrades of beam, FD, and ND are required to achieve physics goals 
%FD 3 and 4 offer opportunity for instrumentation advances and expanded physics scope
%Phase II beam upgrade offers opportunity for more than just neutrinos - broad precision/intensity program at Fermilab
%-Successful implementation of a large international project in the US is good for the whole community

The 2014 P5 report envisioned DUNE/LBNF as an international neutrino facility at the center of US-based particle physics, making world-leading measurements with impact across all of particle physics for decades to come. Observation of CP violation in neutrinos would be an important step in understanding whether leptogenesis is a viable explanation of the baryon asymmetry of the universe. Determination of the neutrino mass ordering will inform design of experiments to determine the Majorana or Dirac nature of neutrinos. Precise measurements of oscillation parameters can confirm the three-flavor model of neutrino oscillation or reveal evidence for additional states or interactions. DUNE will be part of a growing network of experiments for multi-messenger astronomy and neutrinos from a core collapse supernova contain a wealth of information about astrophysical processes as well as particle physics. DUNE's sensitivity to physics beyond the Standard Model offers the ability to significantly constrain models and the potential for paradigm-changing discoveries that are complementary to those at collider experiments and other precision experiments. DUNE's far detector upgrades offer the community the opportunity to develop advanced instrumentation that improves and expands DUNE's physics reach. Measurements of neutrino-nucleus cross sections and other precision electroweak physics at DUNE's near detector will contribute to improved understanding and modeling of nuclear and high-energy physics.
The 2020 Update of the European Particle Physics Strategy reaffirmed the international community's commitment to DUNE, stating that ``Europe, and CERN through the Neutrino Platform ... should continue to collaborate with the United States and other international partners towards the successful implementation of [LBNF] and [DUNE]''\cite{EuropeanStrategyGroup:2020pow}. The current Snowmass/P5 process offers the US community the opportunity to reaffirm and update the exciting vision laid out by P5 and embraced by the international physics community.

DUNE has made significant progress towards realizing the P5 vision: establishment of a large, international scientific collaboration, major investments from the US DOE, international funding agencies, and CERN, sophisticated physics sensitivity analyses, successful prototypes demonstrating the power of the LArTPC detector design, real progress on excavation and conventional facilities, and the start of construction of DUNE far detector components in the UK. As described in Section~\ref{sect:design}, each aspect of the experiment has been specifically designed to meet the physics needs of the experiment. As a result of funding constraints, DUNE has adopted a phased approach to realizing the design; Phase I includes the first half of the required far detector mass, half of the required beam power, and the minimal suite of ND components required to make credible oscillation physics measurements. With this configuration, DUNE will have unique and world-leading ability to definitively determine the neutrino mass ordering, will begin contributing to measurements of oscillation parameters and multi-messenger astronomy, and will begin to gain sensitivity to BSM physics. However,  we emphasize that Phase II, which includes upgrades to at least 40-kt of far detector fiducial mass, a 2.4 MW proton beam, and the full near detector, is required for world-leading study of CP violation, the precision oscillation measurement program, and significant BSM discovery potential. Each of these three upgrades is necessary but not sufficient to exploit DUNE's full potential and DUNE's primary physics program is in significant jeopardy if any of these upgrades are not realized.

The addition of FD-3 and FD-4 is critical for the success of DUNE's accelerator and non-accelerator based physics programs, and offers an exciting opportunity to expand the physics scope of DUNE beyond the original P5 vision. %LArTPC-based 
Far detector module concepts with expanded scope have been proposed, offering the possibility to dramatically expand DUNE's physics reach to include searches for dark matter or neutrinoless double beta decay at the far detector. The community has developed a number of new ideas\cite{mooworkshop}, including a LArTPC concept based on pixel readout~\cite{auger2019new}, one with expanded sensitivity to low-energy physics~\cite{lowbackgroundmodule}, and advanced optical neutrino detectors\cite{Theia:2019non,Lowe:2020wiq}.
DUNE welcomes the possibility of new collaborators and expanded scope from innovations in far detector technology and encourages the community to explore this potential as part of Snowmass. 
% The specific detection technology for these modules does not need to match that of FD-1 and FD-2, but the modules must provide at least equivalent sensitivity to DUNE's primary physics goals.
%The specific detection technology for these modules does not necessarily need to match FD-1 and FD-2; the minimum requirement is simply that they provide at least 10 kt each of fiducial target mass and are otherwise appropriate for the long-baseline oscillation program. Constraints on systematic uncertainties from the ND using the same nuclear target and similar detection principle are critical to DUNE's oscillation physics analysis, so the collaboration has some preference for LArTPC-based detectors.
The specific detection technology for these modules does not need to match that of FD-1 and FD-2, but the modules must provide at least equivalent sensitivity to DUNE’s primary physics goals. Constraints on systematic uncertainties from the ND using the same nuclear target and detection principle as the far detector are critical to DUNE’s oscillation physics sensitivities.
%. One of the key ND constraints comes from the PRISM technique, which is demonstrated to inform FD predictions on LAr.  
Non-LArTPC modules have also been suggested and may be feasible, but a strategy for constraining systematics to the required level must be demonstrated. 

%Do we want to go here? Is it bad to be this direct.
Similarly, the upgrade required to reach 2.4 MW proton beam power will keep Fermilab's accelerator complex at the forefront of the intensity frontier, and offers significant opportunities for new research directions within US particle physics. This upgrade is critical to DUNE's oscillation physics program, but the proton accelerator upgrade that enables the increase in LBNF beam power benefits the entire Fermilab physics program; using it solely for DUNE would be an inefficient use of resources. The improved accelerator complex could support a broad range of experiments including next-generation muon experiments to search for lepton flavor violation, precision measurements of rare decays sensitive to new physics at energy scales well beyond those accessible by direct searches, and fixed-target and beam-dump experiments for direct detection of new physics. DUNE encourages the community to take advantage of the Snowmass process to explore the full range of physics that would be enabled by the beam upgrade.

DUNE's success is critical to the entire US particle physics community, both because of the important and wide-ranging physics potential of the experiment and as a model for future large-scale, US-based international projects. LBNF/DUNE has the potential to facilitate a broad physics program with scope far greater than originally envisioned by P5. DUNE therefore calls on the particle physics community to affirm via the Snowmass process that:
\begin{itemize}
\itemsep0em     
    \item Phase I should be realized within the current decade. Every effort should be made to resolve the funding profile issues that could delay first physics results until the 2030s.
    \item Realization of the full DUNE design is the highest priority. The Phase II upgrades must be aggressively pursued such that Phase II is fully realized within the next decade.
    \item Research to design DUNE detectors with additional capability and to develop other experiments that make use of the beam upgrades should be supported. 
\end{itemize}

%% file: sections/conclusions.tex
\section{Conclusions}
\label{sec:conclusions}

DUNE is designed to make precision measurements of long-baseline neutrino oscillations, as well as to measure low-energy neutrinos, and search for physics beyond the Standard Model. The 2014 P5 report strongly endorsed the DUNE physics program, and laid out a phased approach to realizing the experiment. Significant progress has been made since the P5; the collaboration has developed detailed designs for the near and far detectors, constructed end-to-end physics analysis, and demonstrated the LAr TPC technology in a prototype that is full scale in the drift dimension. DUNE is on track to achieve its Phase I design, consisting of a 1.2 MW neutrino beam, 20kt of LAr fiducial mass at the far detector, and a minimal near detector. This configuration is sufficient to unambiguously determine the neutrino mass ordering, and is sensitive to CP violation for nearly maximal $\deltacp$. However, the long-term measurement program of DUNE requires an upgrade to Phase II: a 2.4 MW beam, 40 kt of LAr fiducial mass, and a more capable near detector. DUNE Phase II is a high-precision neutrino experiment with a broad underground physics program that will make world-leading measurements of neutrino oscillations and searches for new physics over multiple decades. Phase II may include detector innovations that will further broaden the physics reach of DUNE. The Snowmass process is an opportunity for the US particle physics community to re-affirm a commitment to the physics goals described in the last P5 report, to update the P5 vision taking into account the significant advances made in DUNE/LBNF, and to prioritize the full DUNE physics program.

%%%%%%%%%%%%%%%%%%%%%%%%%%%%%%%%%%%%%%%%%%%%%%%%%%%%%%%%%%%

%% file: sections/acknowledgements.tex
\section{Acknowledgements}
\label{sec:acknowledgements}

This document was prepared by the DUNE collaboration using the
resources of the Fermi National Accelerator Laboratory 
(Fermilab), a U.S. Department of Energy, Office of Science, 
HEP User Facility. Fermilab is managed by Fermi Research Alliance, 
LLC (FRA), acting under Contract No. DE-AC02-07CH11359.
%
% Funding agencies, alphabetical by country, then alphabetical by agency name
%
This work was supported by
CNPq,
FAPERJ,
FAPEG and 
FAPESP,                         Brazil;
CFI, 
IPP and 
NSERC,                          Canada;
CERN;
M\v{S}MT,                       Czech Republic;
ERDF, 
H2020-EU and 
MSCA,                           European Union;
CNRS/IN2P3 and
CEA,                            France;
INFN,                           Italy;
FCT,                            Portugal;
NRF,                            South Korea;
CAM, 
Fundaci\'{o}n ``La Caixa'',
Junta de Andaluc\'ia-FEDER,
MICINN, and
Xunta de Galicia,               Spain;
SERI and 
SNSF,                           Switzerland;
T\"UB\.ITAK,                    Turkey;
The Royal Society and 
UKRI/STFC,                      United Kingdom;
DOE and 
NSF,                            United States of America.
%
% Acknowledgement of NERSC if those resources were used
%
This research used resources of the 
National Energy Research Scientific Computing Center (NERSC), 
a U.S. Department of Energy Office of Science User Facility 
operated under Contract No. DE-AC02-05CH11231.
%
%%%%%%%%%%%%%%%%%%%%%%%%%%%%%%%%%%%%%%%%%%%%%%%%%%%%%%%%%%%

%% file: common/final.tex
    % this is added just after end of document

% end stuff from init
%\cleardoublepage
%\renewcommand{\bibname}{References}
\renewcommand{\refname}{References}

%\printglossary

%\bibliographystyle{apsrev4-1}
\bibliographystyle{utphys}
% To understand the style chosen, see:
% https://arxiv.org/hypertex/bibstyles/ (very bottom -- additions) and
% https://www.sharelatex.com/learn/Bibtex_bibliography_styles
% July 2017, AH and BV (and AM)

%\begin{multicols}{2}[\printbibheading]
\bibliography{common/tdr-citedb}
%\printbibliography
%\end{multicols}